\documentclass[11pt]{article}
\usepackage[margin=1in]{geometry}
\usepackage{physics}
\usepackage{amsmath}
\usepackage{amssymb}
\usepackage{amsfonts}
\usepackage{amsthm}
\usepackage{graphicx}
\usepackage[utf8]{inputenc}
\DeclareUnicodeCharacter{2212}{-}

\usepackage{url}

\title{\vspace{-5em}Using physical features of protein core packing to distinguish real proteins from decoys}

\date{\vspace{-5em}}

\usepackage[affil-it]{authblk} 
\usepackage{etoolbox}
\usepackage{lmodern}

\makeatletter
\patchcmd{\@maketitle}{\LARGE \@title}{\fontsize{16}{19.2}\selectfont\@title}{}{}
\makeatother

\usepackage{graphicx}
\usepackage{dcolumn}
\usepackage{bm}

\usepackage[T1]{fontenc}
\usepackage{mathptmx}
\usepackage{gensymb}

\makeatletter
\patchcmd{\@maketitle}{\LARGE \@title}{\fontsize{16}{19.2}\selectfont\@title}{}{}
\makeatother

\title{\vspace{-5em}Current MD forcefields fail to capture key features of protein structure and fluctuations: A case study of cyclophilin A and T4 lysozyme}

\author[1,2]{Zhe Mei}
\affil[1]{Department of Chemistry, Yale University, New Haven, Connecticut, 06520, USA}
\affil[2]{Integrated Graduate Program in Physical and Engineering Biology, Yale University, New Haven, Connecticut, 06520, USA}

\author[2,3]{Alex T. Grigas}
\affil[3]{Graduate Program in Computational Biology and Bioinformatics, Yale University, New Haven, Connecticut, 06520, USA}

\author[2,4]{John D. Treado}
\affil[4]{Department of Mechanical Engineering and Materials Science, Yale University, New Haven, Connecticut, 06520, USA}

\author[5]{Gabriel Melendez Corres}
\affil[5]{Department of Biology, UPR Humacao, Humacao, Puerto Rico 00792}

\author[6]{Maisa Vuorte}
\affil[6]{Department of Chemistry and Materials Science, School of Chemical Engineering, Aalto University, Aalto, Finland}

\author[6,7]{Maria Sammalkorpi}
\affil[7]{Department of Bioproducts and Biosystems, School of Chemical Engineering, Aalto University, Aalto, Finland}

\author[8]{Lynne Regan}
\affil[8]{Institute of Quantitative Biology, Biochemistry, and Biotechnology, Center for Synthetic and Systems Biology, School of Biological Sciences, University of Edinburgh, Edinburgh, UK}

\author[2,9,10]{Zachary A. Levine}
\affil[9]{Department of Pathology, Yale School of Medicine, New Haven, Connecticut, 06520, USA}
\affil[10]{Department of Molecular Biophysics and Biochemistry, Yale University, New Haven, Connecticut, 06520, USA}

\author[2,4,11,12]{Corey S. O'Hern}
\affil[11]{Department of Physics, Yale University, New Haven, Connecticut, 06520, USA}
\affil[12]{Department of Applied Physics, Yale University, New Haven, Connecticut, 06520, USA}

\usepackage[square,numbers,sort&compress]{natbib}
\begin{document}

\maketitle

\textbf{}

\textbf{Abstract:} Globular proteins undergo thermal fluctuations in solution, while maintaining an overall well-defined folded structure.  In particular, studies have shown that the core structure of globular proteins differs in small, but significant ways when they are solved by x-ray crystallography versus solution-based NMR spectroscopy.  Given these discrepancies, it is unclear whether molecular dynamics (MD) simulations can accurately recapitulate protein conformations. We therefore perform extensive MD simulations across multiple force fields and sampling techniques to investigate the degree to which computer simulations can capture the ensemble of conformations observed in experiments. By analyzing fluctuations in the atomic coordinates and core packing, we show that conformations sampled in MD simulations both move away from and sample a larger conformational space than the ensemble of structures observed in NMR experiments. However, we find that adding inter-residue distance restraints that match those obtained via Nuclear Overhauser Effect measurements enables the MD simulations to sample more NMR-like conformations, though significant differences between the core packing features in restrained MD and the NMR ensemble remain. Given that the protein structures obtained from the MD simulations possess smaller and less dense protein cores compared to those solved by NMR, we suggest that future improvements to MD forcefields should aim to increase the packing of hydrophobic residues in protein cores.

\maketitle

\section{Introduction}
\label{sec:level1}

Over the past 50 years, numerous experiments have characterized the three-dimensional structure of globular proteins. For example, there are tens of thousands peptide and protein structures that have been solved by x-ray crystallography with resolutions less than $2$ \AA~\cite{pdb:BermanNucAcidsRes2000,pdb:BurleyNucAcidsRes2018}. In addition, ensembles of structures from more than ten thousand peptides and proteins have been obtained using solution NMR spectroscopy, providing complimentary information to existing crystal structures.

While the application of deep learning methods have improved protein structure predictions~\cite{CASP:MoultProteins2016,CASP:MoultProteins2018,CASP:KryshtafovychProteins2019}, it remains difficult to accurately predict the folded structure of a protein based solely on its sequence. One of the most frequently used methods for 
predicting protein structure and fluctuations is all-atom molecular dynamics
(MD) simulations. Although a large number of peptides have been folded by MD simulations~\cite{peptidefolding:DauraJMB1998,peptidefolding:SchaeferJMB1998,peptidefolding:FerraraJPCB2000,peptidefolding:HoPLoSCompBio2006,peptidefolding:CinoJCTC2012}, only $\lesssim 20$ distinct proteins have been successfully folded starting from non-native conformations using MD simulations. In smaller proteins (with
$N \lesssim 50$ residues), MD folding simulations often recapitulate the
experimental structures with C$_{\alpha}$ root-mean-square deviations (RMSD) less than $1.5$ \AA~between the computational and experimental
structures. However, for larger proteins between $\sim 50$ and $80$ amino acids, predictions from
MD simulations yield backbones that deviate from experimental structures
by more than $3.0$ \AA. (See Appendix~\ref{folding} for a review of recent computational studies of protein folding.)

While prior protein folding simulations have been evaluated across multiple commonly used MD forcefields and for the ability of peptides and proteins to sample specific  conformations, few studies have systematically characterized how fluctuations around the folded state in the MD simulations compare to fluctuations in experimental ensembles. For example, in x-ray crystal structures, one can define an ensemble from proteins that have been crystallized multiple times at high resolution. Such duplicates possess an average backbone ${\rm RMSD} \sim 1.0$ \AA, which is consistent with the backbone RMSD values obtained from B-factors~\cite{subgroup:MeiProteins2020}. Alternatively, thermal fluctuations of proteins in solution can be described by high-quality NMR ensembles, where each model has more than $15$-$20$ restraints per residue. In particular, the intra-bundle backbone RMSD for NMR ensembles plateaus beyond $15$-$20$ restraints per residue at $\sim 1.4$ \AA~\cite{subgroup:MeiProteins2020}. Global backbone fluctuations generated from prior MD simulations of globular proteins (at room temperature) are larger than the fluctuations around the folded state observed in both x-ray crystallography and solution NMR experiments~\cite{subgroup:MeiProteins2020,comp:GarbuzynskiyPSFB2005}. (See Appendix~\ref{folding}.) These previous results emphasize the necessity for a systematic comparison of both local and global fluctuations in protein structure across multiple MD forcefields to determine their ability to recapitulate the fluctuations in the experimentally-derived NMR ensembles.

Prior computational studies have also considered whether experimentally-derived protein structures remain stable or deviate from their experimental conformations when they are used as initial structures in MD simulations across a wide range of atomistic forcefields. For example, in recent work by Robustelli, {\it et al.}, MD simulations initialized with NMR or x-ray crystal structures of small proteins were run for hundreds of $\mu$s, and their $^3J$-couplings, inter-residue distances from Nuclear Overhauser Effect (NOE) measurements, and residual dipolar couplings (RDCs) were compared to experimental values~\cite{shawNMR:RobustelliPNAS2018}. The authors found that the Amber99SB-ILDN forcefield yielded the most stable simulations and the closest agreement with NMR measurements. Thus, one might conclude that MD simulations using contemporary forcefields generally maintain experimentally-determined protein structures. While it is significant that many small proteins do not partially unfold during long MD simulations, this result does not imply that the conformational fluctuations sampled during the MD simulations are accurate, especially for core residues in large proteins. Both the fluctuations of residues on the surface and in the core obtained from MD simulations need to be compared quantitatively to those observed in spectroscopic experiments, such as NMR.

Recently, we showed that the packing of core residues in globular proteins dictates the quality of computational models, and that a failure to correctly pack core residues results in poorly folded model structures~\cite{subgroup:GrigasProSci2020}. Therefore, systematic comparisons of molecular forcefields should also include studies of the important features of protein cores, including their fluctuations. The cores of experimentally-determined protein structures share several key properties, including (1) the fraction of residues that are core (defined by a relative solvent accessible surface area ${\rm rSASA} < 10^{-3}$) is typically between $5-10\%$, (2) the packing fraction $\phi$ (fraction of space occupied by protein atoms) of core residues occurs between $0.54 < \phi < 0.59$, and (3) the atomic overlap among core residues is small, typically less than $0.1$ \AA~per residue. We have previously shown that it is possible to distinguish ‘good’ (with small C$_{\alpha}$ RMSD compared to experimental structures) from ‘bad’ computational models by determining whether the models satisfy the above three core packing properties. These results demonstrate the strong correlation between core structure and the conformation of the entire protein. Furthermore, high-quality protein structures deduced both through x-ray crystallography and NMR spectroscopy reveal that the NMR structures possess higher packing fractions in the core ($\phi \sim 0.59$), even though the total core overlap energy and quality of side chain repacking is the same for x-ray crystal and NMR structures~\cite{subgroup:MeiProteins2020}.  Subtle differences in core properties are important for determining the structure of the entire protein, and therefore should be included when analyzing protein conformations generated by MD simulations.

The prior results discussed above raise an important question. Since high-quality NMR and x-ray crystal structures possess differences in their backbone RMSD and core packing properties, do MD simulations generate protein conformations and fluctuations that are more similar to NMR or x-ray crystal structures?  To address this question, we carry out all-atom MD simulations of two  well-resolved globular proteins with experimentally-determined x-ray and NMR structures, cyclophilin A~\cite{3k0m:FraserNature2009} and T4 lysozyme*~\cite{3dmv:LiuJMB2009}, each containing $N >160$ residues. We evaluate fluctuations in their structure using three commonly used forcefields (CHARMM36m~\cite{charmm36m:HuangNatMet2017}, Amber99SB-ILDN~\cite{a99sb-ildn:BestJPhysChemB2009,a99sb-ildn:Lindorff-LarsenProteins2010}, and Amber99SBNMR-ILDN~\cite{a99sb-nmr:LiAngewandteChemie2010}) to determine whether each can properly recapitulate core packing features of crystal structures or NMR bundles. Amber99SB-ILDN was developed to match secondary structure propensities found in experiments, Amber99SBNMR-ILDN was developed to incorporate NMR measurements, such as chemical shifts and $J$-couplings, and CHARM36m was developed to sample diverse backbone conformations in folded and disordered proteins. We find that the cores of both proteins (in terms of C$_{\alpha}$ RMSD, packing fraction, and fraction of core residues) are far more similar to those found in x-ray crystal structures than those in NMR bundles. Further, all three forcefields fluctuate with a global C$_{\alpha}$ RMSD (relative to the x-ray crystal structure or the NMR bundle) $\sim 3$ \AA, which is larger than the fluctuations that occur in both x-ray crystal structure duplicates and within NMR bundles. By adding restraints among core residues based on NOE measurements from NMR experiments to the MD simulations, we can largely reduce the global C$_{\alpha}$ RMSD (relative to the NMR bundle) in the MD simulations as expected. However, the global C$_{\alpha}$ RMSD is still larger than the value measured from the NMR bundle. Thus, our results indicate that these MD forcefields are unable to capture experimentally measured NMR fluctuations. We also performed replica exchange MD (REMD) simulations to determine if accelerated sampling of the forcefields improves both the sampling of experimentally-determined structures and their fluctuations.  We find that the  structures sampled in the REMD simulations also do not possess the core packing properties of experimentally-determined NMR structures.  

\section{Results}
\label{results}

In this work, we describe all-atom molecular dynamics simulations in explicit solvent of two proteins in an explicit solvent that were initialized with their experimental structures (either an x-ray crystal structure or one of the structures from the NMR bundle) using three of the most commonly used molecular forcefields,  Amber99SB-ILDN, Amber99SBNMR-ILDN, and CHARMM36m.  We chose to study the two globular proteins, cyclophilin A and T$4$ lysozyme*, because they both have been experimentally-determined to high resolution using x-ray crystallography and with a large number of restraints using NMR spectroscopy. (See Sec.~\ref{datasets} for descriptions of the two proteins used in this study.) After initializing the MD simulations with the experimentally-determined structures, we ran long trajectories ($\geq 1 \mu$s), and measured the C$_{\alpha}$ RMSD with respect to the NMR and x-ray crystal structures, the fraction of core residues, their packing fractions, and related structural quantities as a function of time. (As a comparison, the global C$_{\alpha}$ RMSD from previous MD studies of folding and stability are presented in Appendix~\ref{folding}.) The MD simulations were carried out at room temperature, 1 bar of pressure, and in a large cubic box with periodic boundary conditions.  (See Sec.~\ref{methods} for a detailed description of the methodology employed for the MD simulations.)

\begin{figure*}
\begin{center}
\includegraphics[width=0.99\textwidth]{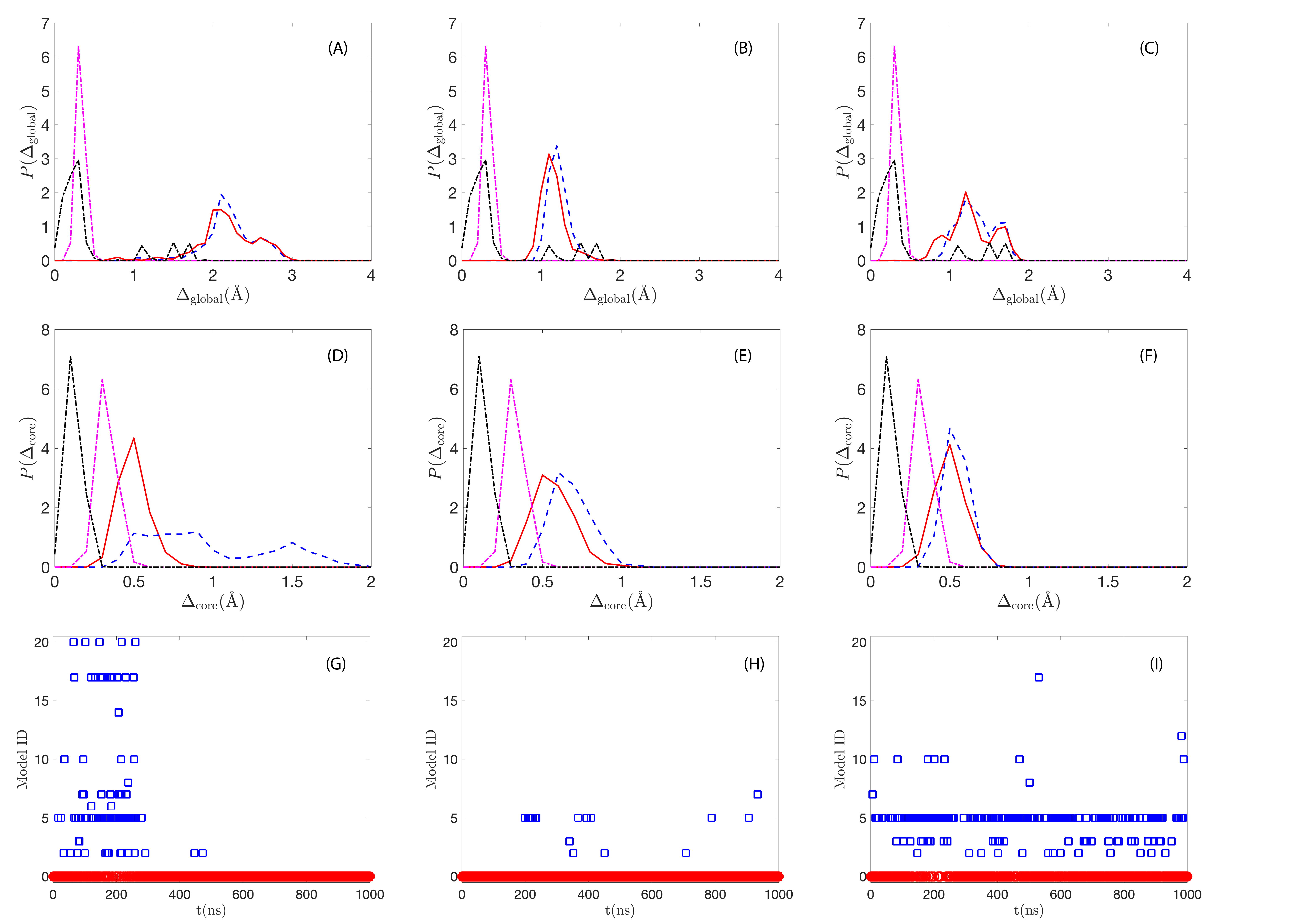}
\caption{Structural fluctuations of cyclophilin A from all-atom MD simulations and experimentally-determined x-ray crystal and solution-based NMR structures. The three columns provide results for the all-atom MD
simulations using the (left) CHARMM36m, (middle) Amber99SB-ILDN, and (right) Amber99SBNMR-ILDN, forcefields. (A)-(C) The probability distributions $P(\Delta_{\rm global})$ of the root-mean-square deviations $\Delta_{\rm global}$ in the positions of the C$_{\alpha}$ atoms of all residues in the protein [in \AA] between structures in the NMR bundle (magenta dot-dashed lines), between the x-ray crystal structure duplicates (black dot-dashed lines), between the structures in the MD simulations and the closest  x-ray duplicates (red solid lines), and between the structures in the MD simulations and closest models in the NMR bundle (blue dashed lines).  (D)-(F)  The probability distributions $P(\Delta_{\rm core})$ of the root-mean-square deviations $\Delta_{\rm core}$ in the positions of the C$_{\alpha}$ atoms of the core residues [in \AA] for the same data in panels (A)-(C). (G)-(I) The identification number of the NMR or x-ray crystal structure with the smallest $\Delta_{\rm core}$ with respect to the MD structure at each time $t$. Model $0$ corresponds to the x-ray crystal structure with PDB code 3k0m, and models $1$-$20$ indicate the NMR models in the bundle ordered from smallest to largest $\Delta_{\rm core}$ relative to the x-ray crystal structure with PDB code 3k0m. The MD simulations in (A)-(I) were initialized using the x-ray crystal structure with PDB code 3k0m.}
\label{fig:RMSD_dist_NMR_Xray}
\end{center}
\end{figure*}

To calibrate the results of the MD simulations, we must first determine the conformational fluctuations in globular proteins in experiments. We previously identified a set of over $20$ proteins that have been characterized multiple times by x-ray crystallography to a resolution of $< 2$ \AA~and by NMR using $> 15$-$20$ restraints per residue~\cite{subgroup:MeiProteins2020}. First, for a given protein, do duplicate x-ray crystal structures and structures from the NMR bundle fluctuate around the same average structure?  Second, is the magnitude of these fluctuations between x-ray duplicates and NMR ensembles the same? We found that, in general, the fluctuations in structure in the NMR ensemble are larger than those for the x-ray duplicate ensemble, and the difference in the average structure between the x-ray and NMR ensembles is larger than the fluctuations within each ensemble separately. For example, for core residues, we found that the average C$_{\alpha}$ RMSD of core residues among x-ray crystal structure duplicates is $\approx 0.1$ \AA, the average C$_{\alpha}$ RMSD of core residues among models within each NMR bundle is $\approx 0.5$ \AA, and the average C$_{\alpha}$ RMSD of core residues between the x-ray duplicates and models in each NMR bundle is $\approx 0.8$ \AA.  (We define core residues as those with sufficiently small relative solvent accessible surface area, ${\rm rSASA}$. See Sec.~\ref{methods} for the definitions of ${\rm rSASA}$ and RMSD.) We also characterized the global fluctuations. We found that the average global C$_{\alpha}$ RMSD among the x-ray crystal structure duplicates is $\approx 0.5$ \AA, the average global C$_{\alpha}$ RMSD among models within each NMR bundle is $\approx 1.2$ \AA, and the average global C$_{\alpha}$ RMSD between the x-ray duplicates and models in each NMR bundle is $\approx 1.8$ \AA.  The differences (both for core residues and globally) between the structures in the x-ray crystal and NMR ensembles are substantially larger than the fluctuations within each ensemble separately. 

Both cyclophilin A and T$4$ lysozyme* are examples from the dataset of proteins that have duplicate high-resolution x-ray crystal structures and high-quality NMR structures. First, we will discuss the results for cyclophilin A. In Fig.~\ref{fig:RMSD_dist_NMR_Xray}, we show that the average core C$_{\alpha}$ RMSD among the x-ray crystal structure duplicates is $\approx 0.1$ \AA~and among models within the NMR bundle is $\approx 0.3$ \AA.
The average global C$_{\alpha}$ RMSD among the x-ray crystal structure duplicates is $\approx 0.4$ \AA~and among the models within the NMR bundle is $\approx 0.5$ \AA. We compare these results to those from MD simulations starting from an x-ray duplicate structure (with PDB code 3k0m).  In Fig.~\ref{fig:RMSD_dist_NMR_Xray} (A)-(C), we show the distributions $P(\Delta_{\rm global})$ of the C$_{\alpha}$ RMSD for all residues in cyclophilin A among x-ray duplicates, among models in the NMR bundle, between structures in the MD simulations and the closest x-ray duplicates, and between structures in the MD simulations and the closest models in the NMR bundle for the three MD forcefields.  For all forcefields, $P(\Delta_{\rm global})$ is shifted to larger values for the MD simulations compared to $P(\Delta_{\rm global})$ for the experimentally-determined x-ray duplicate and NMR ensembles.

\begin{figure*}
\begin{center}
\includegraphics[width=0.66\textwidth]{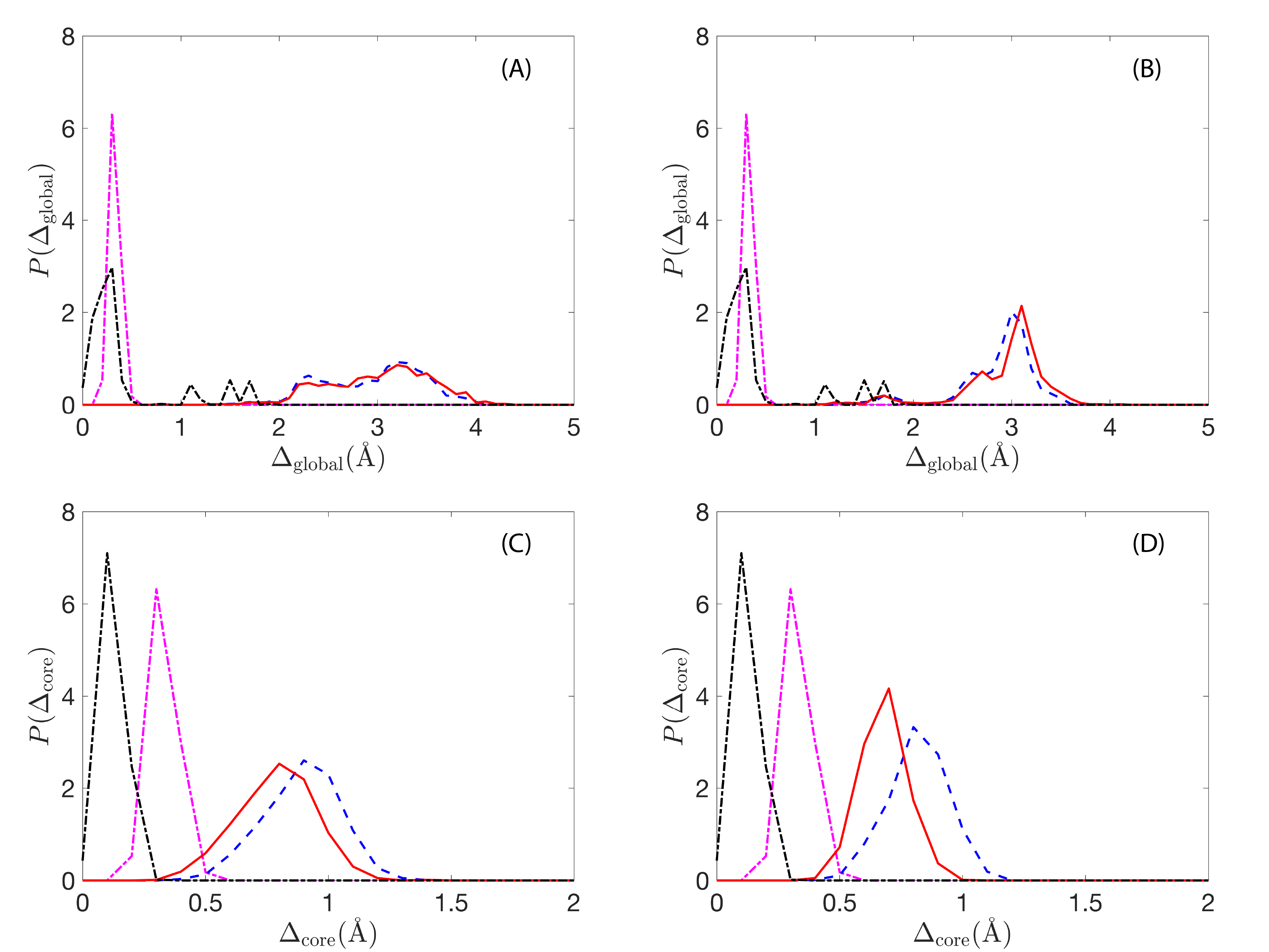}
\caption{(A)-(B) The probability distributions $P(\Delta_{\rm global})$ of the root-mean-square deviations $\Delta_{\rm global}$ in the positions of the C$_{\alpha}$ atoms of all residues in cyclophilin A [in \AA] between structures in the NMR bundle (magenta dot-dashed lines), between the x-ray crystal structure duplicates (black dot-dashed lines), between the structures in the MD simulations (using CHARMM36m) and the closest x-ray duplicates (red solid lines), and between the structures in the MD simulations (using CHARMM36m) and the closest models in the NMR bundle (blue dashed lines). The MD simulations in (A) and (B) were initialized using two different models from the NMR bundle. (C)-(D) The probability distributions $P(\Delta_{\rm core})$ of the root-mean-square deviations $\Delta_{\rm core}$ in the positions of the C$_{\alpha}$ atoms of the core residues [in \AA] for the same data in panels (A) and (B).}
\label{fig:RMSD_initial_structures}
\end{center}
\end{figure*}

Similar results are found for the distribution $P(\Delta_{\rm core})$ of C$_{\alpha}$ RMSD for core residues in Fig.~\ref{fig:RMSD_dist_NMR_Xray} (D)-(F) obtained from MD simulations of cyclophilin A.  For all three forcefields, $P(\Delta_{\rm core})$ is shifted to larger values for the MD simulations compared to $P(\Delta_{\rm core})$ for the experimentally-determined x-ray duplicate and NMR ensembles.  In panel (D), we find that $P(\Delta_{\rm core})$ is particularly broad for the MD simulations with CHARMM36m, extending to $\Delta_{\rm core} > 1.5$ \AA~where there is no weight in $P(\Delta_{\rm core})$ for the experimentally-determined NMR bundle. We find similar results for $P(\Delta_{\rm global})$ and $P(\Delta_{\rm core})$ for the MD simulations when they are initialized using models from the NMR bundle in Fig.~\ref{fig:RMSD_initial_structures}. Similar results are found for MD simulations of cyclophilin A using the two Amber forcefields in Appendix~\ref{initial_conditions}. Thus, for all three forcefields tested in this study, the fluctuations, $\Delta_{\rm global}$ and $\Delta_{\rm core}$, observed in MD simulations for cyclophilin A are in general larger than those observed in the experimentally-determined x-ray duplicate and NMR ensembles. These results show that the MD simulations of cyclophilin A sample a broader set of structures than either the x-ray duplicate or NMR ensembles. 

We also investigated whether the structures sampled in the MD simulations are closer (determined by the smallest C$_{\alpha}$ RMSD, $\Delta_{\rm core}$, for core residues) to a particular structure in the x-ray duplicate or NMR ensembles. In Fig.~\ref{fig:RMSD_dist_NMR_Xray} (G)-(I), we identify the particular structure (either the x-ray crystal structure with PDB code 3k0m, labelled $0$, or one of $20$ models in the NMR bundle, ordered from smallest to largest $\Delta_{\rm core}$ from the x-ray crystal structure) that is closest to the protein structure in the MD simulations as a function of time for each of the three forcefields.  In general, the structures in the MD simulations are closer to the x-ray crystal structure than the models in the NMR bundle. The structures in the MD simulations are closer (minimum C$_{\alpha}$ RMSD) to the x-ray crystal structure $\approx 82\%$, $97\%$, and $56\%$ of the time for CHARMM36m, Amber99SB-ILDN, and Amber99SBNMR-ILDN, respectively.  For the MD simulations with Amber99SBNMR-ILDN, $8$ of the $20$ NMR model structures are sampled. One model is sampled the most, at $20\%$ of the time. 

\begin{figure*}
\begin{center}
\includegraphics[width=0.99\textwidth]{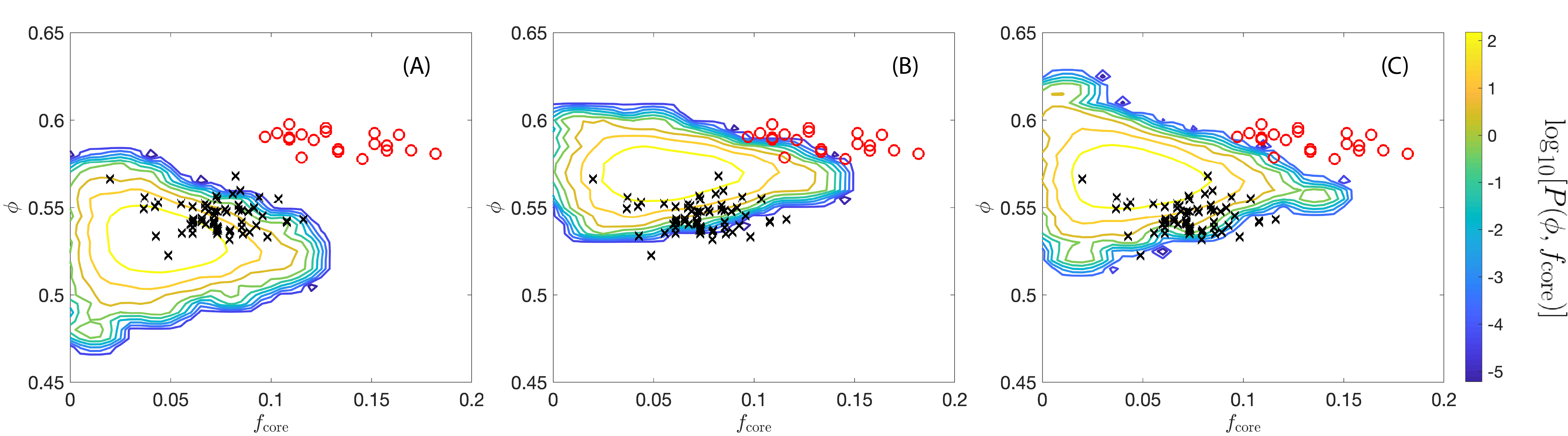}
\caption{Probability distribution $P(\phi,f_{\rm core})$ of the packing fraction $\phi$ of the core and fraction of core resdiues $f_{\rm core}$ from MD simulations of cyclophilin A using the (A) CHARMM36m, (B) Amber99SB-ILDN, and (C) Amber99SBNMR-ILDN forcefields.  The contours are shaded using the color scale from yellow with high probability to dark blue with low probability. (The MD simulations were initialized in the x-ray crystal structure with PDB code 3k0m.) In addition to the results for each forcefield, we include values of $\phi$ and $f_{\rm core}$ for the x-ray crystal structure duplicates (black exes) and all models in the NMR bundle (red open circles).}
\label{fig:phi_vs_fcore}
\end{center}
\end{figure*}

In recent studies, we found that properties of core packing in globular proteins were different for protein structures obtained from x-ray crystallography and from solution NMR spectroscopy.  In particular, we found that the packing fraction of core residues from high-resolution x-ray crystal structures was $\langle \phi \rangle = 0.55 \pm 0.01$, whereas $\langle \phi \rangle = 0.59 \pm 0.02$ for structures obtained from NMR.  Thus, an important question is whether the packing properties of protein cores generated from MD simulations more closely resemble those in x-ray crystal or NMR structures. In Fig.~\ref{fig:phi_vs_fcore} (A)-(C), we show the probability distribution $P(\phi,f_{\rm core})$ for obtaining core packing fraction $\phi$ and fraction of core residues $f_{\rm core} = N_c/N$, where $N_c$ is the number of core residues with ${\rm rSASA} < 10^{-3}$, during MD simulations of cyclophilin A with each of the three forcefields. We compare $P(\phi,f_{\rm core})$ to the $\phi$ and $f_{\rm core}$ values for each structure in the x-ray crystal structure duplicate and NMR dataset for cyclophilin A. As expected, the mean core packing fraction for the NMR structures for cyclophilin A, $\langle \phi \rangle \sim 0.59$, is larger than that for the x-ray crystal structure duplicates, $\langle \phi \rangle \sim 0.54$.  Further, the mean fraction of core resides, $\langle f_{\rm core} \rangle \sim 0.13$, is larger for the NMR structures of cyclophilin A compared to that for the x-ray crystal structure duplicates, $\langle f_{\rm core} \rangle \sim 0.07$.  For the MD simulations with CHARMM36m, the protein samples $\phi$ and $f_{\rm core}$ values that are similar to, but slightly smaller than those for the x-ray crystal structures. For CHARMM36m, the protein never samples the NMR values of $\phi$ and $f_{\rm core}$. For the MD simulations using the Amber99SB-ILDN and Amber99SBNMR-ILDN forcefields, $P(\phi,f_{\rm core})$ shifts to larger values of $\phi$ and $f_{\rm core}$, but few of the NMR-determined values of $\phi$ and $f_{\rm core}$ are sampled. Thus, the MD simulations of cyclophilin A for all three forcefields most frequently sample $\phi$ and $f_{\rm core}$ values associated with the cores of x-ray crystal structures, not the cores of NMR structures. Note that while $P(\phi,f_{\rm core})$ samples x-ray crystal structures more often, the peak of $P(\phi,f_{\rm core})$ is not centered on the data for x-ray crystal structures.

\begin{figure*}
\centering
\includegraphics[width=0.99\textwidth]{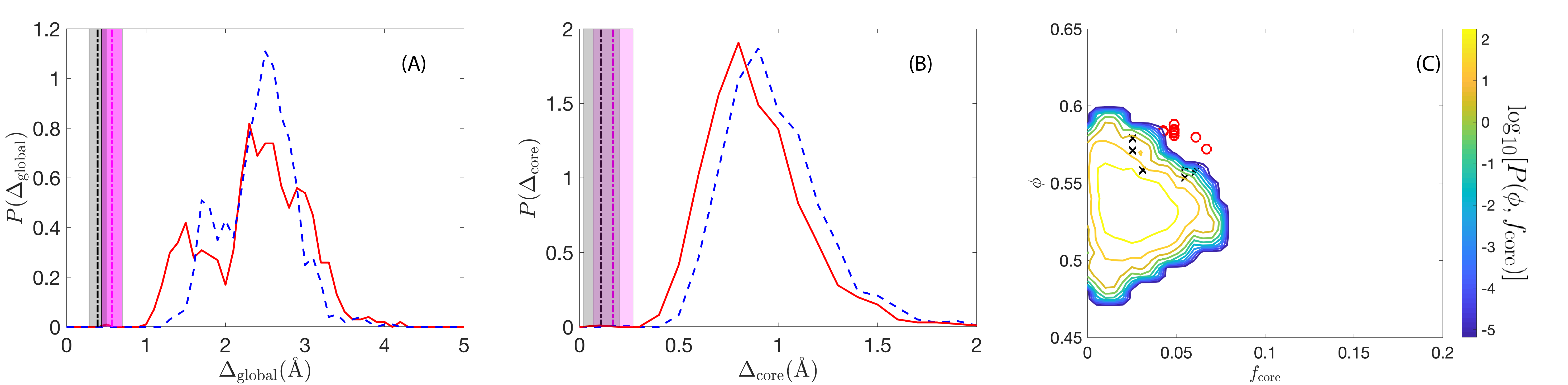}
\caption{(A) Probability distribution $P(\Delta_{\rm global})$ for the C$_{\alpha}$ RMSD for all residues in T4 lysozyme* between the structures generated from MD simulations using CHARMM36m and the closest x-ray duplicates (red solid line) and the closest models in the NMR bundle (blue dashed line).  We also show the mean and standard deviation for $\Delta_{\rm global}$ for the x-ray duplicate dataset (black dashed-dotted line indicates the mean and gray shaded region indicates the standard deviation) and NMR bundle (magenta dashed-dotted line indicates the mean and magenta shaded region indicates the standard deviation). In (B), we show results for $P(\Delta_{\rm core})$ for the same data in (A). We also show the mean and standard deviation for $\Delta_{\rm core}$ for the x-ray duplicate dataset (black dashed-dotted line indicates the mean and gray shaded region indicates the standard deviation) and NMR bundle (magenta dashed-dotted line indicates the mean and magenta shaded region indicates the standard deviation). (C) Probability distribution $P(\phi,f_{\rm core})$ of the packing fraction $\phi$ of the core and fraction of core residues $f_{\rm core}$ from MD simulations of T4 lysozyme* using CHARMM36m.  The contours are shaded using the color scale from yellow with high probability to dark blue with low probability on a logarithmic scale. We also include values of $\phi$ and $f_{\rm core}$ for the x-ray duplicates (black crosses) and all models in the NMR bundle (red open circles). (The MD simulations for T4 lysozyme* were initialized in the x-ray crystal structure with PDB code 3dmv.) }
\label{fig:T4}
\end{figure*}

For comparison, we also performed ($\geq 1\mu$s) MD simulations of T4 lysozyme* using the three forcefields and both the x-ray crystal and NMR structures as initial conditions. The results presented in Fig.~\ref{fig:T4} for MD simulations of T4 lysozyme* are qualitatively similar to those presented in Figs.~\ref{fig:RMSD_dist_NMR_Xray}-\ref{fig:phi_vs_fcore} for cyclophilin A. 
In particular, the core and global C$_{\alpha}$ RMSD between the structures in the MD simulations and the experimentally-determined x-ray crystal and NMR structures are much larger than the respective RMSD measures for the x-ray duplicate and NMR ensembles 
separately. For example, as shown in Fig.~\ref{fig:T4} (A), $\langle \Delta_{\rm global} \rangle \sim 2.7$\AA~between structures in the MD simulations and the closest x-ray duplicates. In contrast, $\langle \Delta_{\rm global} \rangle = 0.4$\AA~for x-ray duplicate structures.  Moreover, as shown in Fig.~\ref{fig:T4} (B), $\langle \Delta_{\rm core} \rangle \sim 0.8$\AA~between structures in the MD simulations and the closest x-ray duplicates, whereas $\langle \Delta_{\rm global} \rangle = 0.2$\AA~for x-ray duplicate structures for T4 lysozyme*. In Fig.~\ref{fig:T4} (C), we show the probability distribution $P(\phi,f_{\rm core})$ from the MD simulations 
of T4 lysozyme*. Although the $\phi$ and $f_{\rm core}$ values for the x-ray crystal and NMR structures are closer together for T4 lysozyme* than for cyclophilin A, we still find that the MD simulations frequently sample smaller values of $\phi$ and $f_{\rm core}$ than those found for the NMR structures. 

In the studies described above, we started the MD simulations with experimentally-determined NMR and x-ray crystal structures as the initial conditions, ran the MD simulations for $1 \mu$s, and showed that multiple MD forcefields sample conformations that are closer to the x-ray crystal structures compared to NMR structures. In addition, all three forcefields failed to sample cores that are as large and densely packed as those in the NMR bundle. In Appendix~\ref{time_dependence}, we describe tests of convergence of the RMSD and radius of gyration as a function of time during the MD simulations. We show that the $1\mu$s MD simulations with the Amber forcefields do not change with time beyond $\sim 100$ns. However, for the MD simulations of cyclophilin A using the CHARMM36m forcefield, the C$_{\alpha}$ RMSD and radius of gyration increase with time, and thus we stopped the MD simulations with CHARMM36m at $1\mu$s and instead used an enhanced sampling technique to explore the forcefield further.

To determine whether the differences in structure and fluctuations between the MD simulations and NMR experiments are, at least in part, due to under-sampling of experimental conformations, we enhanced the sampling of the MD simulations using a replica exchange protocol. If increased computational sampling moves the MD simulations closer to the NMR ensemble, under-sampling is likely occurring. However, if conformations continue to diverge from the NMR ensemble, it is likely that the forcefield possesses low-lying energy minima that are distinct from those in the experimental ensemble. Thus, we performed replica exchange molecular dynamics (REMD) simulations of cyclophilin A to further explore low-energy protein conformations that are sampled by the CHARMM36m forcefield. For the REMD simulations, we considered a partially unfolded initial structure that was prepared at elevated temperatures with $R_g/R_g^0 >2$ (where $R_g^0$ is the radius of gyration of the x-ray crystal structure) and core C$_{\alpha}$ RMSD $\Delta_{\rm core} \approx 1.5$\AA~compared to the x-ray crystal structure and nearest NMR structure, which is much larger than core C$_{\alpha}$ RMSD values among the x-ray crystal structure duplicates or NMR bundle. Initializing the REMD simulations with a partially unfolded structure allows us to determine whether the REMD simulations can find the correct folded structure of cyclophilin A when they are started in a protein structure from a region of conformation space that is distant from the experimentally-determined structures. (See Sec.~\ref{methods} for a detailed discussion of the REMD simulation methodology.)

\begin{figure*}
\begin{center}
\includegraphics[width=0.9\textwidth]{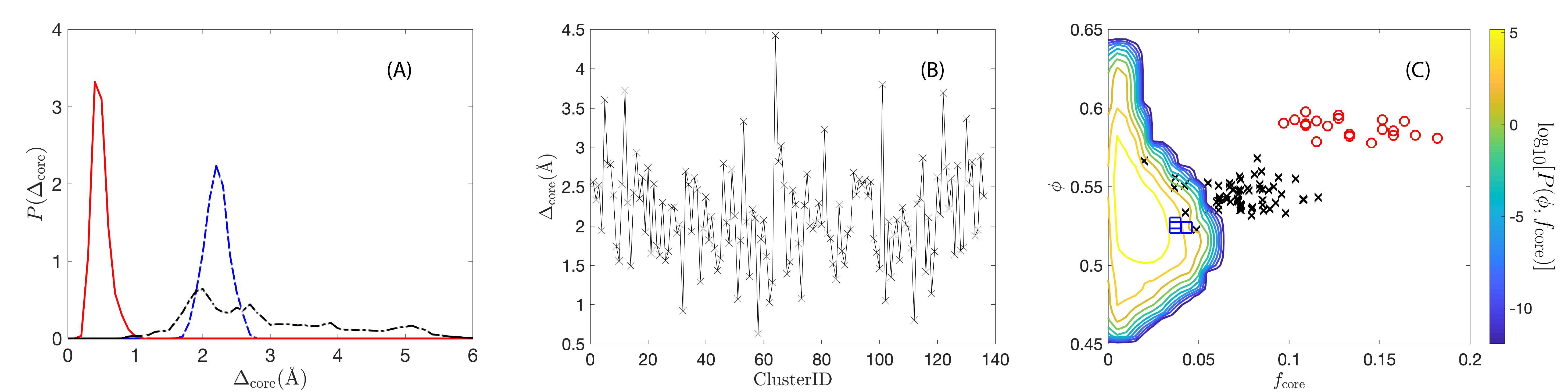}
\caption{(A) Probability distributions of the core C$_{\alpha}$ RMSD $\Delta_{\rm core}$ (relative to the x-ray crystal structure PDB code 3k0m) from MD simulations (at room temperature using CHARMM36m) of cyclophilin A starting from the x-ray crystal structure with PDB code 3k0m (red solid line) and starting from the x-ray crystal structure heated to a temperature above the unfolding temperature (blue dashed line). We also show $\Delta_{\rm core}$ from REMD simulations (at room temperature) starting from the same structure prepared at elevated temperature (black dot-dashed line). (B) The average core C$_{\alpha}$ RMSD $\langle \Delta_{\rm core} \rangle$ (relative to the closest NMR structure) obtained from REMD simulations of cyclophilin A at room temperature starting from the structure prepared at elevated temperature.  The REMD structures obtained from $10^4$ snapshots separated by $200$ ns were grouped into $136$ clusters for which any two structures possess $\Delta_{\rm core} <1$\AA. (C) Probability distribution $P(\phi,f_{\rm core})$ of the packing fraction $\phi$ of the core and fraction of core residues $f_{\rm core}$ from the REMD simulations described in (A) and (B).  The contours are shaded using the color scale from yellow with high probability to dark blue with low probability on a logarithmic scale. We also include values of $\phi$ and $f_{\rm core}$ for the x-ray crystal structure duplicates (black crosses) and all models in the NMR bundle (red open circles) for cyclophilin A. The three open squares indicate the average values of $\phi$ and $f_{\rm core}$ for clusters $31$, $57$, and $111$ from (B).}
\label{fig:remd}
\end{center}
\end{figure*}

We show in Fig.~\ref{fig:remd} (A) that MD simulations started from the partially unfolded structure remain far from the experimental x-ray crystal structure. In particular, the distribution $P(\Delta_{\rm core})$ of core C$_{\alpha}$ RMSD values (relative to the x-ray crystal structure PDB: 3k0m) for the MD simulations starting from the partially unfolded structure is peaked at $\Delta_{\rm core} \sim 2.2$\AA, whereas it is peaked at $\Delta_{\rm core} \sim 0.5$\AA~when the starting structure is the x-ray crystal structure with PDB code 3k0m. In contrast, when we initialize the REMD simulations with the partially unfolded structure, $P(\Delta_{\rm core})$ is broad, sampling structures over a range of $\Delta_{\rm core}$ from $0.75$\AA~to greater than $5$\AA~(relative to the x-ray crystal structure). Thus, REMD simulations, even though they are initialized with partially unfolded structures, are able to sample, albeit infrequently, structures that are close to (less than $1$\AA) the x-ray crystal structure. Thus, the REMD protocol might be promising for protein structure prediction. However, one must be able to identify when the REMD simulation is close to the experimental structure, without knowing the experimental structure beforehand. The distance to the x-ray crystal structure can be estimated using protein decoy detection methods~\cite{subgroup:GrigasProSci2020,ProQ3:UzielaScientificReports2016,3dcnn:SatoPLoS2019,SBROD:Karasikovbioinformatics2018}, which have their own limitations, and would limit the ability of REMD to identify experimental conformations. It is likely that the CHARMM36m forcefield possesses low-lying energy minima that are distinct from those sampled by x-ray crystal structures. 

We also investigate whether these REMD simulations started from a partially unfolded structure sample conformations that are close to the NMR structures for cyclophilin A. In Fig.~\ref{fig:remd} (B), we first divide the REMD simulation ($10^4$ frames) into $136$ clusters, where each pair of structures within each cluster has $\Delta_{\rm core} < 1$\AA. We then calculate the average core C$_{\alpha}$ RMSD $\Delta_{\rm core}$ between the structures in each cluster and the nearest structure in the NMR bundle. In Fig.~\ref{fig:remd} (B), we plot $\Delta_{\rm core}$ versus the cluster label, where the clusters are ordered based on the number of REMD frames in each. Overall, most of the REMD clusters have large $\Delta_{\rm core}$ compared to the closest structure in the NMR bundle. However, three clusters (with cluster labels $31$, $57$, and $111$) possess $\Delta_{\rm core} < 1$\AA~relative to the closest structure in the NMR bundle. In Fig.~\ref{fig:remd} (C), we plot the distribution $P(f_{\rm core}, \phi)$ for the REMD simulations starting from a partially unfolded structure and find that overall the packing properties of the structures in the REMD simulations are different than those found in the x-ray crystal and NMR structures.  In fact, $P(f_{\rm core}, \phi)$ samples regions of large packing fraction $0.58 < \phi <0.65$ and small fraction of core residues $0 < f_{\rm core} < 0.025$ that were not sampled in the MD simulations of cyclophilin A initialized with the experimentally-determined structures.  It is possible that the REMD simulations, which are seeded with conformations at higher temperatures, bias the system towards higher packing fractions as has been observed in MD simulations of jammed packings of amino acid-shaped particles~\cite{subgroup:MeiProteins2020}. The average packing fraction of core residues $\phi$ and fraction of core residues $f_{\rm core}$ for clusters $31$, $57$, and $111$ are closer to the values for x-ray crystal structures than for the NMR models. Even after exploring a broad region of conformation space, most favorable conformations in the REMD simulations do not possess core packing features that are similar to those in the NMR bundle.

We have demonstrated via several metrics (local and global C$_{\alpha}$ RMSD, fraction of core residues, and packing fraction) that current MD forcefields (CHARMM36m, Amber99SB-ILD and Amber99SBNMR-ILDN) fail to recapitulate protein structure fluctuations in NMR bundles. One possible method for accurately simulating protein structure fluctuations would be to add harmonic restraints between atom pairs for which we have NOE measurements and other NMR data. For example, in the Biological Magnetic Resonance Bank (BMRB)~\cite{bmrb:UlrichNucAcidsRes2007}, we find that there are a total of $4101$ NOE atom pairs and $127$ pairs involving atoms in core residues for which NOE measurements have been performed for cyclophilin A. (All of the restraints between core heavy atoms, $24$ pairs, are listed in Appendix~\ref{restraints}.) By adding harmonic restraints for the $127$ atomic separations involving core residues (using spring constants that are comparable to those for covalently bonded atoms), we are able to reduce $\Delta_{\rm core}$ and $\Delta_{\rm global}$ as shown in Table~\ref{table:average}.  (See Sec.~\ref{methods} for a description of the implementation of the harmonic restraints between atom pairs in the MD simulations.) The NMR bundle satisfies all of its core NOE restraints, as the models were fit to this data. The unrestrained MD simulations using the CHARMM36m forcefield only satisfy $49$\% of the core NOE restraints on average. The MD simulations with harmonic restraints using the CHARMM36m forcefield recapitulate $\approx 73\%$ of the core NOE restraints.  Thus, other competing forces in the CHARMM36m forcefield prevent the remaining atomic pair separations from satisfying the NOE restraints. There are at least two ways to improve the frequency with which the MD simulations satisfy the NOE distance measurements: increase the spring constant for the harmonic restraints among core NOE atom pairs or increase the number of restraints, for example, by including harmonic restraints between non-core and core NOE atom pairs. We did not increase the spring constant of the harmonic restraints above the values of carbon-carbon bonds because this can lead to unphysical stretching of covalents bonds. Adding more restraints yields an MD simulation methodology that must be tailored for each individual protein, and is not applicable to a broad set of globular proteins. We do not believe this is a fruitful approach.

\begin{figure*}
\begin{center}
\includegraphics[width=0.9\textwidth]{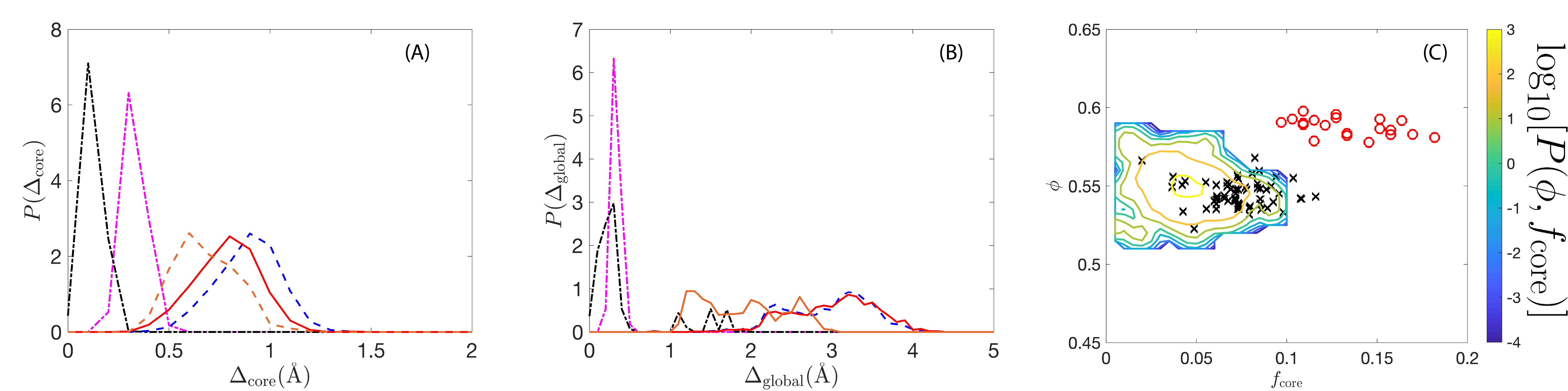}
\caption{The probability distributions of the (A) global $\Delta_{\rm global}$ and (B) core RMSD $\Delta_{\rm core}$ of C$_{\alpha}$ atoms between structures obtained from the restrained MD simulations of cyclophilin A (using CHARMM36m) compared to the NMR bundle (orange dashed line). We also show $\Delta_{\rm global}$ and $\Delta_{\rm core}$ for the x-ray crystal structure duplicates (black dot-dashed line), NMR bundle (magenta dot-shed line), and for structures from the unrestrained MD simulations of cyclophilin A (CHARMM36m) compared to the NMR bundle (blue dotted line) and x-ray crystal structure duplicates (red solid line). (C) Probability distribution $P(\phi,f_{\rm core})$ of the packing fraction $\phi$ of the core and fraction of core residues $f_{\rm core}$ from restrained MD simulations of cyclophilin A using CHARMM36m. We also show values of $\phi$ and $f_{\rm core}$ from x-ray crystal structure duplicates (black crosses) and all models in the NMR bundle (red open circles). The contours are shaded using the color scale from yellow with high probability to dark blue with low probability on a logarithmic scale. (The MD simulations were initialized in the x-ray crystal structure with PDB: 3k0m.) }
\label{fig:figure6}
\end{center}
\end{figure*}

In Fig.~\ref{fig:figure6} (A) and (B), we show that the core and global C$_{\alpha}$ RMSD distributions for the restrained MD simulations (orange dotted lines) are shifted to lower values compared to those for the unrestrained MD simulations for cyclophilin A. However, the RMSD distributions for the restrained simulations still yield larger RMSD values than those sampled by the structures in the NMR bundle. In addition, we find in Fig.~\ref{fig:figure6} (C) that the packing fractions sampled by the restrained MD simulations are shifted upward relative to values for the unrestrained MD simulations (Fig.~\ref{fig:phi_vs_fcore} (A)), but the fractions of the residues that are identified as core remains low, $f_{\rm core} < 0.1$, whereas $0.1 < f_{\rm core} < 0.18$ for structures sampled by the NMR ensemble. While restraining core atomic separations can reduce the fluctuations of cyclophilin A, only constraining the core atomic separations is not sufficient for recapitulating the conformational fluctuations of cyclophilin A. The non-core region of the protein becomes less dense than the experimental structure and exposes the core region, which leads to low values of $f_{\rm core}$.  Future studies are needed to determine the number and type of atomic distance restraints that are needed to maintain the core properties of the NMR ensemble of structures for cyclophilin A.

\section{Discussion and Conclusions}
\label{conclusions}

We have seen in our previous studies that high-quality structures obtained from x-ray crystallography and NMR spectroscopy possess different distributions of the C$_{\alpha}$ RMSD for core residues (e.g. $\langle \Delta_{\rm core} \rangle$ is larger for the NMR ensemble compared to that for x-ray crystal structure duplicates) and different core packing fractions and core sizes~\cite{subgroup:MeiProteins2020}. Possible explanations for these differences include the experimental conditions. For example, the crystalline environment and typical low temperatures used in x-ray scattering studies of proteins are different from the conditions for solution-based NMR spectroscopy carried out at room temperature. An important goal of MD simulations is to understand the stability of x-ray crystal structures when they are used as initial conditions in MD simulations with explicit solvent at room temperature. For example, do MD simulations initialized with x-ray crystal structures and run with explicit solvent at room temperature yield structures similar to those in the NMR ensemble or do they remain close to the x-ray crystal structure?  

The goal of this article was to address this question using several quantities that characterize protein structure. We conducted long molecular dynamics simulations of two large proteins starting in different experimentally-determined structures using three commonly used forecfields. We found that the RMSD fluctuations of backbone C$_{\alpha}$ atoms in the core and globally in the MD simulations to be both different in magnitude and character compared to the fluctuations in experimentally-determined structures. The conformations sampled in MD simulations are also closer to the x-ray crystal structures than the NMR structures, even when we use structures from the NMR bundle as initial conditions. Additionally, both the size and packing fraction of the cores generated in the MD simulations are more similar to those in x-ray crystal structures, while also sampling many conformations that are more solvent exposed than in x-ray structures. Overall, the MD simulations of these two proteins create smaller and less densely packed cores than those found for structures in the NMR ensemble. Further sampling of CHARMM36m, via REMD simulations, did not generate conformations that are more similar to x-ray crystal or NMR structures. Finally, we showed that by adding harmonic NOE atomic distance restraints, we can reduce the core and global C$_{\alpha}$ RMSD relative to the experimentally-determined structures, although further investigation is needed to identify a minimal set of atomic distance restraints that are needed to recapitulate the core structure.

\begin{figure*}
\begin{center}
\includegraphics[width=0.9\textwidth]{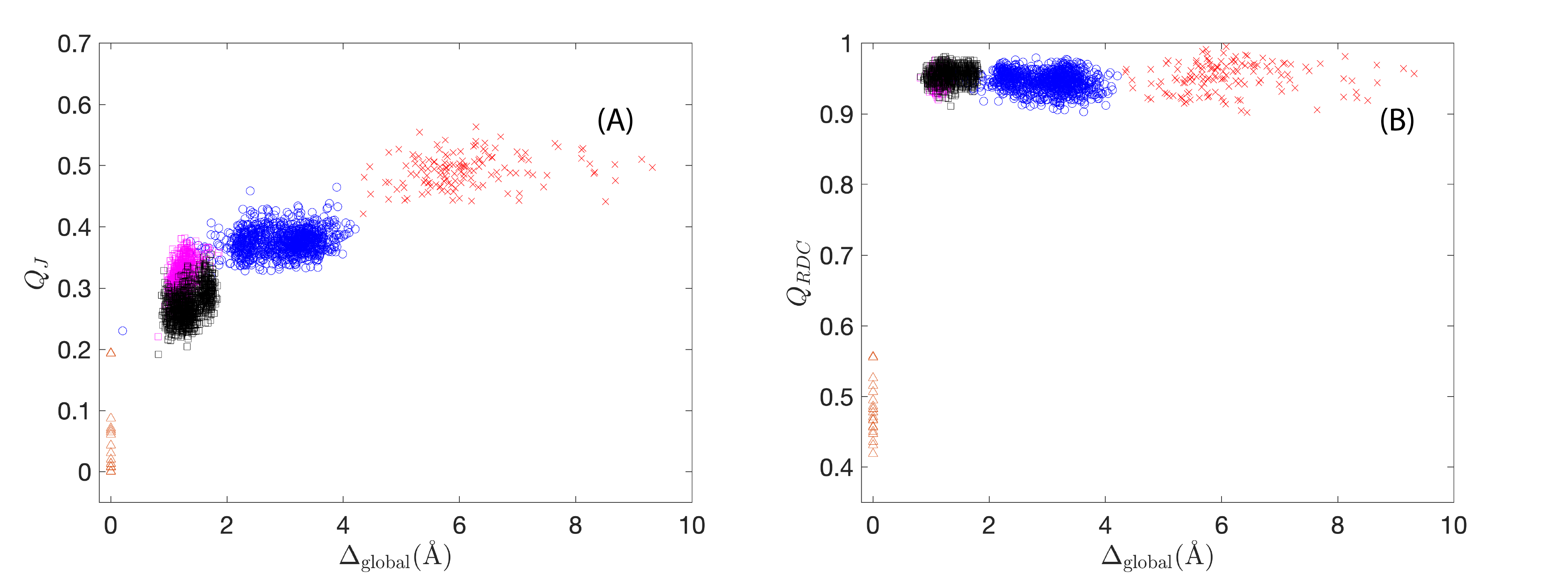}
\caption{The deviation $Q$ in Eq.~\ref{eq:Q} of the (A) $^3J$-coupling and (B) RDC values measured in NMR experiments and in MD simulations of cyclophilin A plotted versus the global C$_{\alpha}$ RMSD $\Delta_{\rm global}$ between the structures from the MD simulations and the closest structure in the NMR bundle. Data from the NMR bundle are plotted at $\Delta_{\rm global}=0$ and shown as orange triangles, and the MD simulations are indicated by black squares (Amber99SB-ILDN), magenta squares (Amber99SBNMR-ILDN), and blue circles (CHARMM36m), and the REMD simulations are represented by red crosses.}
\label{fig:nmrcomp}
\end{center}
\end{figure*}

In the Results section, we compared the atomic coordinates for the structures generated from the MD simulations and the experimental structures in the NMR ensemble. The atomic coordinates in the NMR bundle are obtained through a process of successively incorporating NMR measurements of scalar couplings, chemical shifts, residual dipolar couplings (RDC), NOE atomic separations, and others as restraints on the set of atomic coordinates. Other computational studies of protein folding and stability have compared their MD simulation results to primary NMR experimental data~\cite{stability:LangeBPJ2010,nmrcomp:BestBPJ2008,nmrcomp:BeauchampJCTC2012,shawNMR:RobustelliPNAS2018,shawNMR:Lindorff-LarsenPloSOne}. (Note that even when comparing MD simulations to NMR measurements, one typically uses approximate classical methods to convert atomic coordinates into NMR measurements~\cite{PALES:Zweckstetter2008}.) To compare our MD simulations to NMR measurements, we calculated the deviation $Q_J$ of the $^3J$-couplings obtained from the MD simulations of cyclophilin A relative to the values from the NMR bundle. (See Fig.~\ref{fig:nmrcomp} (A) and Eq.~\ref{eq:Q} in Sec.~\ref{methods}.) Similar to our results above, the two Amber forcefields show smaller $Q_J$ values for $^3J$-couplings than those for the CHARMM36m forcefield. This result agrees with previous MD simulation studies of smaller proteins~\cite{shawNMR:RobustelliPNAS2018}. However, the $Q_J$ values from the MD simulations are at least a factor of $3$ larger than those from the NMR measurements. 

RDC data provides longer-range spatial information than $^3J$-couplings and, unlike NOE measurements, which give atomic separations, RDC data also provides information on the relative orientations of the bond vectors. For cyclophilin A, the $^3J$-couplings were used to determine the structures in the NMR bundle, whereas the RDC data was made available after the NMR bundle was released.  We used the Prediction of Alignment from Structure (PALES) software to calculate the RDC values from the atomic coordinates of the NMR bundle and MD simulations~\cite{PALES:Zweckstetter2008}. The RDC data agrees with the structures in the NMR bundle (i.e. $Q_{\rm RDC} \sim 0.48$~\cite{RDCQ:CornilescuJACS1998,RDCfluc:OttigerJBioNMR1999}), but not as closely as the $^3J$-couplings do ($Q_{J} \sim 0.05$)~\cite{RDCData:CamilloniPNAS2014}. In Fig.~\ref{fig:nmrcomp} (B), we show that the structures in the MD simulations possess $Q_{\rm RDC}$ values between $0.9$ and $1.0$, whereas $\langle Q_{\rm RDC} \rangle \sim 0.48$ for the NMR bundle. The $Q_{\rm RDC}$ values from the MD simulations are much larger than those previously reported for simulations of smaller proteins using similar forcefields~\cite{RDCfluc:ShowalterJACS2007,shawNMR:Lindorff-LarsenPloSOne,shawNMR:RobustelliPNAS2018}. By applying restraints on all measured $^3J$-couplings and NOE atomic separations in cyclophilin A, restrained MD simulations using a forcefield similar to Amber99SB-ILDN can sample conformations with $Q_{\rm RDC} \sim 0.3$~\cite{RDCData:CamilloniPNAS2014}. In contrast, the unrestrained MD simulations of cyclophilin A possess large $Q_{\rm RDC}$ values.  Fig.~\ref{fig:nmrcomp} (B) further emphasizes the sensitivity of RDC values in MD simulations, since $Q_{\rm RDC}$ grows rapidly with $\Delta_{\rm global}$.

Overall, we have seen across three state-of-the-art forcefields, CHARMM36m, Amber99SB-ILDN, and Amber99SBNMR-ILDN, that MD simulations of two large proteins starting from NMR structures, x-ray crystal structures, and a partially unfolded structure, do not adequately recapitulate numerous important properties of the NMR ensemble.  In particular, the cores sampled in the MD simulations are smaller and less densely packed than structures in the NMR bundle. Not a single conformation sampled in the MD simulations captured features in the space of core packing fraction and fraction of core residues sampled by the structures in the NMR bundle. A possible method to improve current MD forcefields is to more accurately model protein hydrophobic cores. For example, the strength of the van der Waals attraction between atoms on hydrophobic residues can be increased and the atomic sizes can be changed to those of the hard-sphere plus stereochemical constraint model. We have shown in previous studies that this model is sufficient to recapitulate side chain conformations of core residues in globular proteins~\cite{subgroup:GainesPED2017}.

\begin{table}[h]
\centering
\begin{tabular}{lllll}
\hline
                        & $\langle \Delta_{\rm core}\rangle$ & $\langle \Delta_{\rm global}\rangle$ & $\langle f_{\rm NOE}\rangle$ \\ \hline
NMR Bundle       & $0.30$                                       & $0.50$                                         & $1.0$                                         \\
No restraints       & $1.00$                                       & $2.50$                                         & $0.49\pm0.11$                                         \\
Core restraints & $0.75$                                       & $1.50$                                         & $0.73\pm0.13$
\\ \hline
\end{tabular}
\caption{The first row indicates the average core $\langle \Delta_{\rm core} \rangle$ and global C$_{\alpha}$ RMSD $\langle \Delta_{\rm global} \rangle$ relative to the closest NMR structure and the fraction of the NOE distance restraints $f_{\rm NOE}$ that are satisfied for structures in the NMR bundle for cylclophilin A.  The second and third rows provide $\langle \Delta_{\rm core} \rangle$ and $\langle \Delta_{\rm global} \rangle$ from MD simulations of cyclophilin A (using the CHARMM36m forcefield) with and without NOE pairwise distance restraints between atoms belonging to core residues.  For the restrained and unrestrained MD simulations of cyclophilin A, we also show the average fraction of NOE pairwise atomic separations that are satisfied for each snapshot. The error bars give the standard deviation of $f_{\rm NOE}$ from the average of $10^4$ snapshots.}
\label{table:average}
\end{table}

\section{Methods}
\label{methods}

\subsection{MD simulations}
\label{md}

We used the GROMACS molecular dynamics package to carry out all of the MD and REMD simulations~\cite{gromacs:AbrahamSoftwareX2015}. For cyclophilin A, the initial structures were obtained from PDB: 1oca~\cite{1oca:OttigerJMB1997} (NMR ensemble) and PDB: 3k0m~\cite{3k0m:FraserNature2009} (x-ray crystal structure). For T4 lysozyme*, the initial structures were obtained from PDB: 3dmv (NMR ensemble) and PDB: 2lcb (x-ray crystal structure). The proteins were solvated with TIP3P water molecules for all forcefields. Short-range van der Waals and screened Coulomb interactions were truncated at $1.2$ nm, while longer-ranged electrostatics were tabulated using the Particle Mesh Ewald summation method. For cyclophilin A, the x-ray crystal structure (PDB: 3k0m) was solved at a pH of $7.5$, whereas the NMR structure was solved at pH $6.5$. For T4 lysozyme*, the pH of the x-ray crystal structure (PDB:2lcb) was $6.9$ and for the NMR structure, the pH was 5.5. All MD simulations in this study were performed at a pH of 7.

Two rounds of energy minimization were performed before the production simulations. The first energy minimization relaxed the water molecules, while fixing the positions of the atoms in the protein, and the second energy minimization relaxed both the protein atoms and water molecules. In Table~\ref{table:EM}, we show how the energy minimization (using the CHARMM36m forcefield) changes the initial structure of cyclophilin A.  We find that energy minimization moves the NMR models by $\Delta_{\rm core} \sim 0.5$\AA. In contrast, energy minimization of the x-ray crystal structure only gives rise to $\Delta_{\rm core} \sim 0.02$\AA. Also, if we apply energy minimization to an NMR structure and calculate $\Delta_{\rm core}$ with respect to the x-ray crystal structure, the deviation is smaller than the deviation relative to the NMR structure. Thus, energy minimization moves NMR structures toward the x-ray crystal structure. These results suggest that x-ray structures are more stable than NMR structures in the CHARMM36m forcefield\cite{charmm36m:HuangNatMet2017}. We find qualitatively similar results to those in Table~\ref{table:EM} for the two Amber forcefields, Amber99SB-ILDN~\cite{a99sb-ildn:BestJPhysChemB2009,a99sb-ildn:Lindorff-LarsenProteins2010} and Amber99SBNMR-ILDN~\cite{a99sb-nmr:LiAngewandteChemie2010}. 

After energy minimization, we performed the MD simulations in the NPT ensemble at temperature $300$K and $1$ bar of isotropic pressure using the weakly coupled Berendsen thermostat and barostat, with a box size that is twice the crystal unit cell on average in each dimension and cubic periodic boundary conditions in all directions. The time constant of the Berendsen thermostat was set to $2$ ps and the isothermal compressibility for the Berendsen barostat was set to $4.5\times10^{-5} \rm bar^{-1}$. The equations of motion for the atomic coordinates and velocities were integrated using a leapforg algorithm with a $2$ fs time step. The simulations were run for $1 \mu$s and sampled every $100$ ps. Tests for convergence of the RMSD and radius of gyration with time are discussed in Appendix~\ref{time_dependence}.  

To better sample conformation space, we also carried out replica exchange MD (REMD) simulations~\cite{remd:SugitaChemPhysLet1999}. The REMD simulations were performed at constant NVT with the volume that gives $P=1$ bar at $300$K using a Nos\'{e}-Hoover thermostat, with a time constant of $1$ ps, a leapfrog integration algorithm, and a time step of $2$fs.  While the MD simulations were carried out in the NPT ensemble, NPT REMD is not implemented in the current version of GROMACS. In future studies, we will compare REMD  sampling in both the NVT and NPT ensembles~\cite{remd:OkabeChemPhysLet2001,remd:MalolepszaJACSB2015,remd:YamauchiJCP2017}. Replicas of the seed systems were duplicated and heated to temperatures ranging from $270$K to $500$K. The ensemble contained $89$ replicas in total, such that the Markovian exchange rate between the replicas was fixed at $25\%$. REMD simulations were run on average for $300$ns per replica. The first $100$ns of the trajectories were ignored to allow for equilibration, and the following $200$ ns period was analyzed.

We also performed restrained MD simulations of cyclophilin A by adding harmonic constraints between NOE atom pairs coupled with the CHARMM36m forcefield. We first identified the NOE restraints for cyclophilin A in the Biological Magnetic Resonance Data Bank (BMRB)~\cite{bmrb:UlrichNucAcidsRes2007}. We found that there are $4101$ restraints in total and $127$ restraints between atoms in core residues. For the restrained simulations, we added ``pseudo-bonds" using a flat-bottom pair potential (i.e. type $10$ restraints) $V_{dr}$ between core atoms with NOE restraints.  No force acts on the atom pair when its separation is within the flat region of the potential, however, a harmonic restoring force acts on the pair to move them into the flat region when the separation is outside of the flat region. The pair potential is given by the following:
\begin{equation} 
\label{eq:potential}
V_{dr}(r_ {ij})  = \begin{cases}
\frac{k_{dr}}{2}(r_{ij}-r_{0})^2, & r_{ij}<r_0\\
0, & r_{0} \leq r_{ij} < r_1\\
\frac{k_{dr}}{2}(r_{ij} - r_1)^2, & r_1 \leq r_{ij} < r_2\\
\frac{k_{dr}}{2}(r_2-r_1)(2r_{ij}-r_2 - r_1), & r_2 \leq r_{ij},\\
\end{cases}
\end{equation}
where $r_{ij}$ is the separation between NOE atom pairs, $r_0$ and $r_1$ are the minimum and maximum distances for the NOE atom pairs in the NMR bundle, $r_2$ is the upper bound of the atom pair separation provided in the BMRB restraint file, and the spring constant $k_{dr}=10$kJ/mol/nm$^2$. The spring constant is on the same order of magnitude as that for bonded heavy atoms pairs.
With the restraints, we first performed energy minimization, then $2$ ps of NVT equilibration, followed by an NPT production run for $1\mu$s. As for the unrestrained MD simulations, we maintained the temperature at $300$K and $1$ bar of isotropic pressure using the weakly coupled Berendsen barostat and thermostat. 

\begin{table*}
\centering
\begin{tabular}{ccccc}
Initial Condition  & Global/Core   & $\Delta(i_{\rm EM},j_0)$ & $\Delta(i_{\rm EM},j_{\rm xray})$  & $\Delta(i_{\rm EM},j_{\rm NMR})$ \\ \hline
 NMR   & Global & $0.81$\AA & $0.21$\AA & $0.79$\AA \\
  & Core & $0.47$\AA & $0.13$\AA & $0.43$\AA  \\
x-ray & Global   & $0.10$\AA & $0.10$\AA & $0.55$\AA \\
  & Core      & $0.02$\AA & $0.02$\AA & $0.39$\AA \\
\end{tabular}
\caption{The root-mean-square deviations (RMSD) $\Delta(i,j)$ in the positions of the C$_{\alpha}$ atoms between two cyclophilin A structures: $i$ and $j$. The first column indicates whether the simulation was initialized with one of the models from the NMR bundle or the x-ray crystal structure PDB: 3k0m. The second column indicates whether the C$_{\alpha}$ RMSD is calculated over core residues or all residues in the protein.  Three RMSD calculations were performed and displayed in the third, fourth, and fifth columns: between the energy minimized ${i}_{\rm EM}$ and initial structures ${j}_0$, between the energy minimized ${i}_{\rm EM}$ and the x-ray crystal structure ${j}_{\rm xray}$, and the energy minimized structure ${i}_{\rm EM}$ and the structures in the NMR bundle ${j}_{\rm NMR}$.}
\label{table:EM}
\end{table*}

\subsection{Datasets}
\label{datasets}

We selected cyclophilin A from an NMR/x-ray pair dataset, which we constructed in our previous work and contains 21 proteins that have been solved by both x-ray crystallography and solution-NMR spectroscopy~\cite{subgroup:MeiProteins2020}. The NMR bundle for cyclophilin A was solved by Ottiger, {\it et al.}~\cite{1oca:OttigerJMB1997} (PDB: 1oca). There are $20$ model structures in the bundle and all of their core residues have more than $20$ NOE restraints. The x-ray crystal structure of cyclophilin A was solved by Fraser, {\it et al.}~\cite{3k0m:FraserNature2009} (PDB: 3k0m) with a resolution of $1.25$ \AA.  We also queried the PDB and found $31$ duplicate x-ray crystal structures of cyclophilin A. These structures have resolution $<2$\AA~and the sequence similarity is greater than $95\%$ compared to the structure with PDB:3k0m. To consider the generality of our results for cylcophilin A, we also performed MD simulations of T4 lysozyme* (which is a mutant of T4 lysozyme with four point mutations, R12G, C55T, C97A, and I137R). This particular mutant has both high quality x-ray and NMR structures. The x-ray crystal structure for T4 lysozyme* was solved by Liu, {\it et al.}~\cite{3dmv:LiuJMB2009} (PDB:3dmv) at resolution $1.65$ \AA~and the NMR ensemble was determined by Bouvignies, {\it et al.}~\cite{2lcb:BouvigniesNature2011} (PDB:2lcb) with $6$ model structures. We also identified $9$ x-ray crystal duplicate structures for T4 lysozyme* with resolution $<2$\AA~and sequence similarity $> 95\%$ compared to the x-ray crystal structure with PDB:3dmv. 

For the analyses of the MD and REMD simulations, each protein conformation was pre-processed using the REDUCE software package, which sets the bond lengths for C-H, N-H, and S-H to $1.1$, $1.0$, and $1.3$ \AA~and the bond angles to $109.5^{\degree}$ and $120^{\degree}$ for hydrogen bond angles involving the C$_{\rm{sp}3}$ and C$_{\rm{sp}2}$ atoms, respectively~\cite{reduce:WordJMB1999}. We also set the values for the atomic radii to be the following: C$_{\rm sp 3}$:$1.5$\AA; C$_{\rm O}$: $1.3$\AA; O: $1.4$\AA; N:$1.3$\AA; H$_{\rm C}$:$1.10$\AA; H$_{\rm O,N}$:$1.00$\AA, and S:$1.75$\AA, which were obtained in prior work by minimizing the difference between the side-chain dihedral angle distributions predicted by the hard-sphere dipeptide mimetic model and those observed in protein crystal structures for a subset of amino acid types~\cite{subgroup:ZhouBPJ2012}. Using these atomic radii, we quantified the relative solvent accessible surface area and the number and packing fraction of core residues.

\subsection{Root-mean-square deviations}
\label{rms_calc}

We measured the root-mean-square deviation (RMSD) of the C$_{\alpha}$ atom positions between two structures $i$ and $j$ after alignment:
\begin{equation}
\label{eq:rmsd}
\Delta (i,j) = \sqrt{\frac{1}{\rm N_s}\sum_{\mu = 1}^{\rm N_s}(\vec c_{\mu, j} - \vec c_{\mu, i})^2},
\end{equation}
where $\vec c_{\mu,j}$ is the position of the C$_{\alpha}$ atom on residue $\mu$ in structure $i$, and $N_{\rm s}$ is the total number of residues that are being compared on the two structures. 

We calculated both the core and global C$_{\alpha}$ RMSD ($\Delta_{\rm core}$ and $\Delta_{\rm global}$) between structure pairs. To calculate the core RMSD, we used the core residues to both align and then compute the RMSD between the two structures. (Identification of core residues is discussed in Sec.~\ref{r_SASA_calc}.) Between the experimental structure pairs (NMR/NMR,x-ray/x-ray, and NMR/x-ray) the set of core residues varies by $\lesssim 20\%$, therefore, we use the union of core residues in both structures. When an MD conformation is compared to an experimental structure, we use the set of core residues defined in the experimental structure. When calculating global RMSD, we align the structures excluding the first and last $4$ amino acids. In addition, when we report the RMSD between an MD conformation and structures in the NMR bundle,  we first calculate the RMSD between the MD conformation and each structure in the NMR bundle, as all structures in the NMR bundle are equally valid. Then, we report the minimum RMSD in this set. The Biopython package was used to align the structures and calculate the RMSD~\cite{biopython:BioInfo2003,biopython:BioInfo2009}.

\subsubsection{Relative solvent accessible surface area}
\label{r_SASA_calc}

To identify core residues, we measured each residue's solvent accessible surface area (SASA). To calculate SASA, we use the NACCESS software package~\cite{naccess:Hubbard1993}, which implements an algorithm originally proposed by Lee and Richards~\cite{rsasa:LeeJMB1971}. The algorithm takes z-slices of the protein, determines the solvent accessibility of the sets of contours using a probe molecule of a given radius, and integrates the SASA over the slices. We use a water-molecule-sized probe with radius 1.4 \AA~and z-slices with thickness $\Delta z = 10^{-3}$\AA, which were used in previous work~\cite{subgroup:GainesPRE2016,subgroup:GainesPED2017,subgroup:TreadoPRE2019,subgroup:MeiProteins2020,subgroup:GrigasProSci2020}. To normalize the SASA, we take the ratio of the SASA within the context of the protein (SASA$_{\rm context}$) and the SASA of the same residue $X$ extracted from the protein structure as a dipeptide (Gly-X-Gly) with the same backbone and side-chain dihedral angles:
\begin{equation}    
\label{rsasa_eq}
{\rm rSASA} = \frac{\rm SASA_{\rm context}}{\rm SASA _{\rm dipeptide}}.
\end{equation}
Core residues are classified as those that have $\rm rSASA \leq 10^{-3}$. This is the largest value of rSASA such that the packing fraction and side-chain repacking predictability no longer depend on the value of the rSASA cutoff when it is decreased.

\subsubsection{Packing fraction}
\label{packing}

A characteristic measure of the packing efficiency of a system is the dimensionless volume fraction, or packing fraction. The packing fraction of residue $\mu$ is
\begin{equation}
\label{packing_frac_eq}
\phi_{\mu} = \frac{v_{\mu}}{V_{\mu}},
\end{equation}
where $v_{\mu}$ is the non-overlapping volume of residue $\mu$ and $V_{\mu}$ is the volume of the surface Voronoi volume of residue $\mu$. To calculate the Voronoi tessellation for a given protein core, we employ surface Voronoi tessellation~\cite{surfacevoro:FabianPhilMag2013}, which defines a Voronoi cell as the region of space that is closer to the bounding surface of residue $\mu$ than to the bounding surface of any other residue in the protein. We calculate the surface Voronoi tessellation using POMELO software package~\cite{pomelo:WeisEPJ2017}. This software approximates the bounding surfaces of each residue by triangulating points on the residue surfaces. We find that using $\sim 400$ points per atom, or $\sim 6400$ surface points per residue, gives an accurate representation of the surface Voronoi cells and the results do not change if more surface points are included.

\subsection{Comparison of NMR structures to those generated by MD simulations}
\label{comparison}

We compared the conformations generated from the MD simulations to NMR measurements in several ways. In particular, we computed $^3J$-coupling constants using the Karplus equation and we calculated residual dipolar couplings (RDCs) using the  PALES software~\cite{PALES:Zweckstetter2008}. $^3J$-coupling constants are determined by the dihedral angle among the two coupled nuclei and the two heavy atoms on either side. For example, the backbone dihedral angle $\phi$ among the backbone atoms C-N-C$_{\alpha}$-C determines $^3J_{H_{N}H_{C_\alpha}}$. To compute the $^3J_{H_N H_{C_{\alpha}}}$ coupling constants from the MD simulations, we used the Karplus equation:
\begin{equation}
\label{karplus}
^3J_{H_{N}H_{C_\alpha}}(\phi) = A\cos^2(\phi + \Theta) + B\cos(\phi + \Theta) + C,
\end{equation}
where $\Theta$ is the phase shift, and $A$, $B$, and $C$ are constants. We used the parameterization of Hu and Bax~\cite{karplusparams:HuJACS1997}: $A = 7.09$ Hz, $B = -1.42$ Hz, $C = 1.55$ Hz, and $\Theta = -60^{\circ}$. Although there are other parameter sets, the correlation between MD simulation and experimental $^3J$ couplings are insensitive to the specific choice of the Karplus parameters~\cite{karplusparams:HuJACS1997}. Also, a single set of Karplus parameters can be applied equally well to all residue types.  

To compare the experimental and calculated values of the NMR measurements, we computed the deviation:
\begin{equation}
Q_S = \frac{\sqrt{N^{-1}_m \sum_{i=1}^{N_m} (S^{\rm calc}_i-S^{\rm exp}_i)^2}}{\sqrt{N^{-1}_m \sum_{i=1}^{N_m} (S^{\rm exp}_i)^2}},
\label{eq:Q}
\end{equation}
where the sum is over the $N_m$ available measurements of the quantity $S$ (i.e. either the $^3J$-couplings or RDCs), $S^{\rm exp}_i$ is the experimental value, and $S^{\rm calc}_i$ is calculated from the structure's atomic coordinates~\cite{RDCQ:CornilescuJACS1998}.

\section*{Acknowledgments}
The authors acknowledge support from NIH Training Grant No. T32GM008283 (A.T.G.) and T32EB019941 (J.D.T.), the Integrated Graduate Program in Physical and Engineering Biology (Z.M.), NSF Grant Nos. DBI-1755494 (G.M.C.), PHY-1522467 (C.S.O), and PHY-2012406 (C.S.O), as well as Fortum Foundation (M.V.), Academy of Finland Grant No. 309324 (M.S.), CSC-IT Centre for Science Ltd. Finland (M.V.). Computational resources were provided by an XSEDE research allocation award No. MCB190047. This work also benefited from the facilities and staff of the Yale University Faculty of Arts and Sciences High Performance Computing Center. 

\section*{Data Availability}
The data that support the findings of this study are available from the corresponding author upon reasonable request.

\section*{Appendix A: MD simulations of cyclophilin A with different initial conditions}
\label{initial_conditions}

\begin{figure*}
\begin{center}
\includegraphics[width=0.6\textwidth]{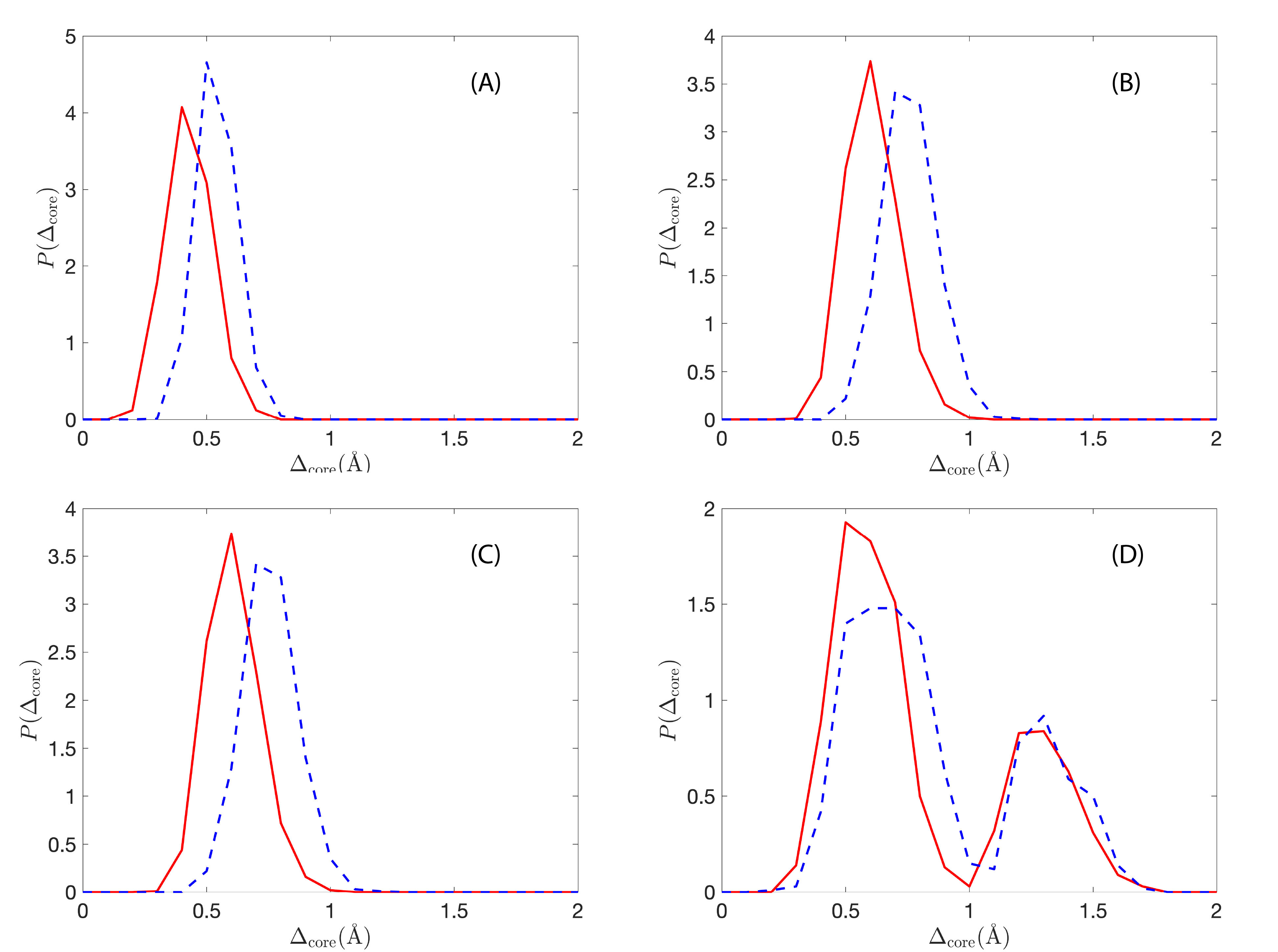}
\caption{Probability distributions of the core C$_{\alpha}$ RMSD $P(\Delta_{\rm core})$ from MD simulations of cyclophilin A using the (A) Amber99SBNMR-ILDN forcefield and (B) Amber99SB-ILDN forcefields starting from the x-ray crystal structure PDB: 3k0m.  $P(\Delta_{\rm core})$ from MD simulations of cyclophilin A using the (C) Amber99SBNMR-ILDN and (D) Amber99SB-ILDN forcefields starting from one of the structures in the NMR bundle (PDB code: 1oca). In all panels, the core C$_{\alpha}$ RMSD is calculated relative to the NMR structure that gives the minimum $\Delta_{\rm core}$ (blue dashed lines) or relative to the x-ray crystal structure PDB: 3k0m (red solid lines). (See Fig.~\ref{fig:RMSD_initial_structures} (C)-(D) for similar results using the CHARMM36m forcefield.}
\label{fig:initial_conditions}
\end{center}
\end{figure*}

In this Appendix, we show that the core RMSD from MD simulations of cyclophilin A initiated from multiple structures in the NMR bundle are smaller when calculated relative to the x-ray crystal structure PDB: 3k0m compared to when calculated relative to the NMR bundle.  In Fig.~\ref{fig:RMSD_initial_structures} (C)-(D), we showed this result for the CHARMM36m forcefield.  In Fig.~\ref{fig:initial_conditions}, we show similar results for the Amber99SB-ILDN and Amber99SBNMR-ILDN forcefields. 

\section*{Appendix B: Testing convergence of MD simulations}
\label{time_dependence}

\begin{figure*}
\begin{center}
\includegraphics[width=0.8\textwidth]{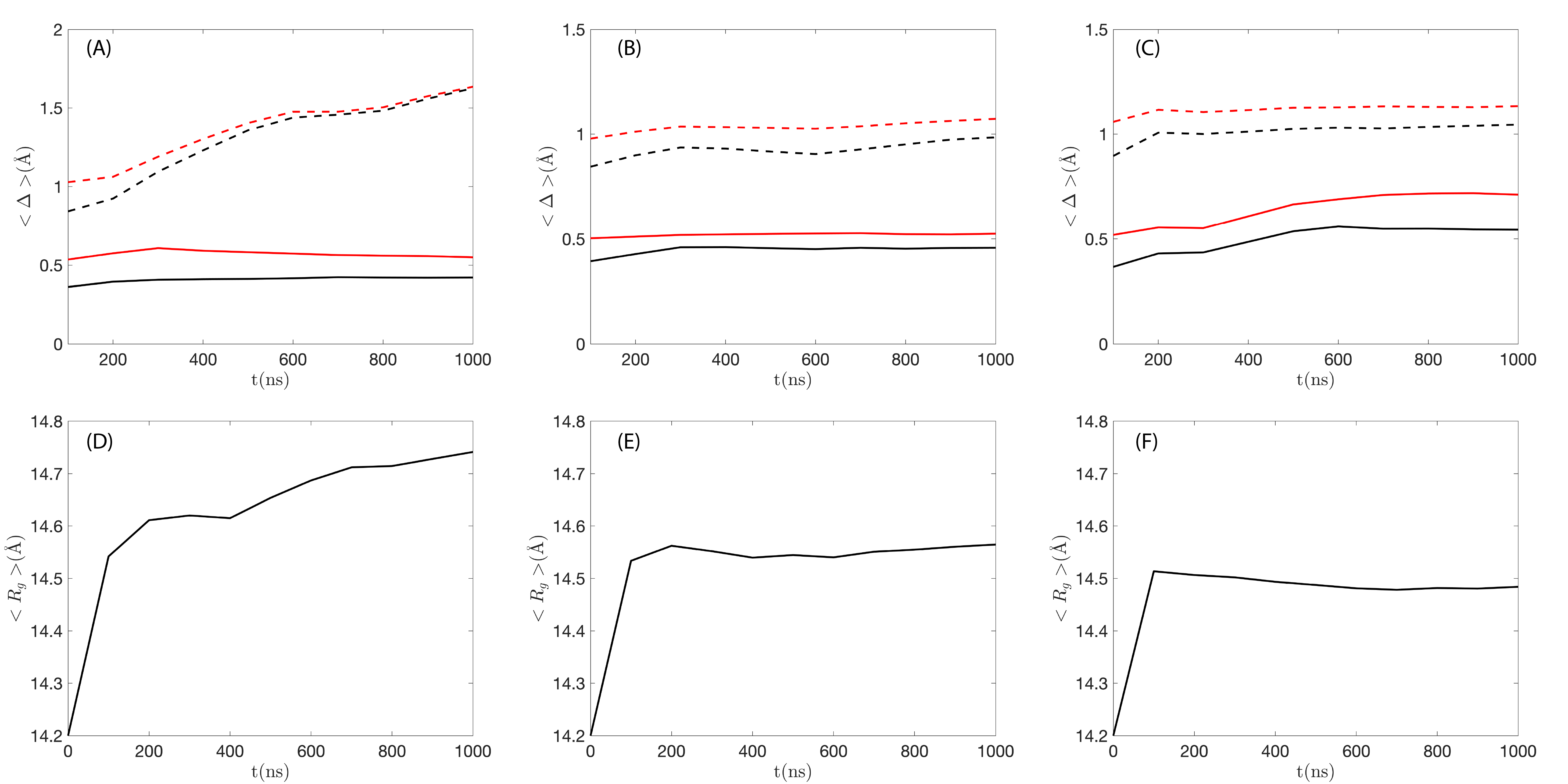}
\caption{The core and global C$_{\alpha}$ RMSD, $\Delta_{\rm core}$ (solid lines) and $\Delta_{\rm global}$ (dotted lines) as a function of time $t$ from MD simulations of cyclophilin A averaged over frames from $0$ to $t$ for the (A) CHARMM36m, (B) Amber99SB-ILDN, and (C) Amber99SBNMR-ILDN forcefields. The RMSD of the protein conformations are calculated relative to NMR bundle (red lines) and x-ray crystal structure (black lines). Similar data is also shown for the average radius of gyration $R_{g}$ as a function of time for (D) CHARMM36m, (E) Amber99SB-ILDN, and (F) Amber99SBNMR-ILDN.}
\label{fig:time_dependence}
\end{center}
\end{figure*}

To assess the convergence of the MD simulations, we calculated the average core and global C$_{\alpha}$ RMSD, $\langle \Delta_{\rm core} \rangle$ and $\langle \Delta_{\rm global} \rangle$, as well as the average radius of gyration $\langle R_{g} \rangle$, for cyclophilin A as a function of time. For the two Amber forcefields, both $\langle \Delta_{\rm core} \rangle$ and $\langle \Delta_{\rm global} \rangle$ plateau after $\sim 100$ns as shown in Fig.~\ref{fig:time_dependence} (B) and (C). In contrast, for the CHARMM36m forcefield, the core RMSD plateaus, but $\langle \Delta_{\rm global} \rangle$ continues to increase beyond $1000$ns. (See Fig.~\ref{fig:time_dependence} (A).) We find the same results for $\langle R_{g} \rangle$ versus time in Fig.~\ref{fig:time_dependence} (D)-(F), indicating that longer-time MD simulations of cyclophilin A using CHARMM36m will lead to partial unfolding, whereas the MD simulations using the two Amber forcefields are stationary in time. As a result, we do not carry out MD simulations of cyclophilin A longer than $1\mu$s, and instead use REMD simulations to explore additional conformations for cyclophilin A. 

\section*{Appendix C: MD simulations of protein folding and stability}
\label{folding}

In this Appendix, we provide a summary of all-atom MD simulations of globular protein folding starting from non-native conformations and globular protein stability starting from experimentally-determined structures. Overall, most folding simulations have been conducted on small proteins containing less than $80$ residues. Simulations of protein stability have been performed on larger proteins, but none larger than the proteins considered in the present studies.

\subsection{Folding simulations}
\label{folding_section}

We identified $15$ globular proteins that have been folded from their primary structures using all-atom MD simulations as shown in Table~\ref{table:folding}. These proteins range from $10$-$80$ amino acids and they all have relatively short folding times ($ < 1\mu$s). Proteins with $\lesssim 35$ residues have been folded within $\Delta_{\rm global} \sim 1$\AA ~of the experimentally-determined structure. For larger proteins, $\Delta_{\rm global}$ begins to increase, reaching $\sim 5$\AA~for some proteins with more than $55$ residues.  In contrast, $\Delta_{\rm global} \sim 1$-$2$\AA~for high-quality NMR bundles~\cite{subgroup:MeiProteins2020}.

\begin{table}[]
\centering
\begin{tabular}{lllllll}
\hline
Protein       & $N$ & $\Delta_{\rm global}$(\AA) & Expts & Forcefields & Reference                                                                                    \\ \hline
Chignolin     & $10$          & $1.0$                                     & NMR   & C22         & \cite{folding:DeSHAWSCIENCE2011}                                            \\
CLN025        & $10$          & $0.9$                                     & NMR   & A11         & \begin{tabular}[c]{@{}l@{}}\cite{folding:PANGReview2017135}\end{tabular} \\
Trp-Cage      & $20$          & $1.4$                                     & NMR   & C22         & \cite{folding:DeSHAWSCIENCE2011}                                            \\
Trp-Cage      & $20$          & $1.7$                                     & NMR   & A11         & \cite{folding:PANGReview2017135}                                            \\
BBA           & $28$          & $1.6$                                     & NMR   & C22         & \cite{folding:DeSHAWSCIENCE2011}                                            \\
Villin        & $35$          & $1.3$                                     & x-ray & C22         & \cite{folding:DeSHAWSCIENCE2011}                                            \\
WW-DOMAIN     & $35$          & $1.2$                                     & x-ray & C22         & \cite{folding:DeSHAWSCIENCE2011}                                            \\
NTL9          & $39$          & $0.5$                                     & x-ray & C22         & \cite{folding:DeSHAWSCIENCE2011}                                            \\
NTL9          & $39$          & $3.0$                                     & x-ray & Gromos      & \begin{tabular}[c]{@{}l@{}}\cite{folding:KamenikJCP2020}\end{tabular}    \\
NTL9          & $39$          & $1-2$                                     & x-ray & A99SB-ILDN      & \cite{folding:ChenJPC2015}                                                                            \\
BBL           & $47$          & $1.2$                                     &  NMR & C22         & \cite{folding:DeSHAWSCIENCE2011}                                            \\
Protein B     & $47$          & $3.3$                                     &  NMR     & C22         & \cite{folding:DeSHAWSCIENCE2011}                                                                                            \\
Homeodomain   & $52$          & $3.6$                                     &   NMR    & C22         & \cite{folding:DeSHAWSCIENCE2011}                                            \\
Protein G     & $56$          & $4.8$                                     &  x-ray     & C22         & \cite{folding:DeSHAWSCIENCE2011}                                            \\
A3D           & $73$          & $4.8$                                     &   NMR    & C22         & \cite{folding:DeSHAWSCIENCE2011}                                            \\
$\lambda$-repressor    & $80$          & $1.8$                                     &   x-ray    & C22         & \cite{folding:DeSHAWSCIENCE2011}                                            \\ \hline
\end{tabular}
\caption{Globular proteins longer than $10$ amino acids that have been folded from their primary structure using MD simulations with specified forcefields. The column ``Expts" reports the experimental method used to solve the protein structure and the global C$_{\alpha}$ RMSD of the MD conformations was calculated relative to this type of experimental structure. Amber is abbreviated as ``A" and CHARMM is abbreviated as ``C".}
\label{table:folding}
\end{table}

\subsection{Stability simulations}
\label{folding_stability}

We identified a number of prior MD simulations of protein stability, where the experimentally-determined  structures are used as initial conditions in the MD simulations in explicit solvent at room temperature. These prior studies have characterized the conformational fluctuations in $19$ distinct proteins ranging in size from $48$-$224$ amino acids. Some of these MD simulations are listed in Table~\ref{table:stability} and others are provided in Table S20 in Robustelli, {\it et al.}~\cite{shawNMR:RobustelliPNAS2018}. The range of the total simulation times varies broadly, from $1$ns to $10\mu$s.  The average global C$_{\alpha}$ RMSD $\Delta_{\rm global} > 1$\AA~for all of the MD simulations. In Robustelli, {\it et al.}~\cite{shawNMR:RobustelliPNAS2018}, the authors carried out $20 \mu$s MD simulations for $14$ proteins using six different forcefields. Only one of the $14$ proteins was determined via NMR spectroscopy, the rest were characterized using x-ray crystallography with a resolution $\leq 2.3$\AA. Amber99SB*-ILDN with TIP3P showed the best performance, with an average RMSD over the final $1 \mu$s of each simulation across all 14 proteins of $2.1$\AA.

\begin{table}[h]
\centering
\begin{tabular}{llllll}
\hline
Protein     & $N$ & x-ray/NMR & Forcefield & $\Delta_{\rm global}$(\AA) & Ref. \\
\hline
Crambin & $46$ &  x-ray & C36m & $\sim 0.75$ & \cite{charmm36m:HuangNatMet2017}  \\ \hline
Homeodomain & $52$            & NMR      & ENCAD       & $3.58$         &  \cite{stability:LevittJCTC1995}    \\ \hline
GB3         & $56$            & x-ray    & A99SB-ILDN  & $1.3$        &   \cite{stability:LiJPCB2011}   \\ 
 &  &  & C36m & $\sim 0.9$ & \cite{charmm36m:HuangNatMet2017}  \\ 
 &  &   & OPLS & $1.0$ & \cite{opls:RobertsonJCTC2015}  \\ 
  &  &   & A03 & $\sim 2.0$ & \cite{stability:LangeBPJ2010}  \\ 
    &  &   & A99sb & $\sim 1.75$ & \cite{stability:LangeBPJ2010}  \\ \hline
BPTI & $58$ &  x-ray & C36m & $1.75$ & \cite{charmm36m:HuangNatMet2017}  \\ \hline
Erabutoxin B & $62$ &  x-ray & C36m & $2.0$ & \cite{charmm36m:HuangNatMet2017}  \\ \hline
CSPA & $69$ &  x-ray & C36m & $1.8$ & \cite{charmm36m:HuangNatMet2017}  \\ \hline
Ubiquitin   & $76$            & x-ray      & A99SB-ILDN  & $1.07$  &   \cite{stability:GANOTH2013}   \\ 
 &  &   & OPLS & $1.0$ & \cite{opls:RobertsonJCTC2015}  \\ 
 &  &   & C36m & $\sim 2.0$ & \cite{charmm36m:HuangNatMet2017}  \\ \hline
Apomyoglobin        & $153$            & x-ray    & GAFF  & $2.50$            &   \cite{stability:ZhangSciRep2017}    \\ \hline
RAP74 C-term & $73$ &  x-ray & C36m & $\sim 3.0$ & \cite{charmm36m:HuangNatMet2017}  \\ \hline
Trimmed ribo-S6 & $74$ &  x-ray & C36m & $\sim 1.75$ & \cite{charmm36m:HuangNatMet2017}  \\ \hline
DMAP1 & $75$ &  x-ray & C36m & $\sim 1.0$ & \cite{charmm36m:HuangNatMet2017}  \\ \hline
ICaBP & $75$ &  x-ray & C36m & $\sim 3.5$ & \cite{charmm36m:HuangNatMet2017}  \\ \hline
PDZ & $85$ &  x-ray & C36m & $\sim 1.1$ & \cite{charmm36m:HuangNatMet2017}  \\ \hline
TEL & $129$ &  x-ray & C36m & $\sim 2.0$ & \cite{charmm36m:HuangNatMet2017}  \\ \hline
FABP & $131$ &  x-ray & C36m & $\sim 1.5$ & \cite{charmm36m:HuangNatMet2017}  \\ \hline
DroHb & $153$ &  x-ray & A14SB & $\sim 1.0$ & \cite{stability:PlanaJCIM2019}  \\ \hline
Mb & $153$ &  x-ray & C36m & $\sim 2.5$ & \cite{charmm36m:HuangNatMet2017}  \\ \hline
DTB Syn & $224$ &  x-ray & C36m & $\sim 2.0$ & \cite{charmm36m:HuangNatMet2017}  \\ \hline

\end{tabular}
\caption{MD simulations of proteins for which the experimental structures (with the method given in the third column) are used as the initial conditions. The global C$_{\alpha}$ RMSD of the conformations in the MD simulations are calculated relative to the experimental structure indicated in the third column. Amber is abbreviated as ``A" and CHARMM is abbreviated as ``C."}
\label{table:stability}
\end{table}

\section*{Appendix D: Restrained simulations}
\label{restraints}

To carry out the restrained MD simulations, 
we obtained information about the NOE restraints from the BRMD website, https://bmrb.io. In the restrained MD simulations, we applied harmonic constraints between $127$ pairs of core atoms. In this Appendix, we list $23$ of the atom pairs, i.e. those that involve core heavy atoms, in Table~\ref{table:restraints}.

\begin{table}[h]
\centering
\begin{tabular}{llllll}
\hline
ResID & Res Type & Atom & Res ID & Res Type & Atom \\ \hline
6     & VAL      & CG1  & 22     & PHE      & CD1  \\
6     & VAL      & CG1  & 22     & PHE      & CD2  \\
6     & VAL      & CG2  & 22     & PHE      & CD1  \\
6     & VAL      & CG2  & 22     & PHE      & CE1  \\
6     & VAL      & CG2  & 24     & LEU      & CD2  \\
6     & VAL      & CG2  & 98     & LEU      & CD1  \\
22    & PHE      & CD1  & 98     & LEU      & CD1  \\
22    & PHE      & CD1  & 98     & LEU      & CD2  \\
22    & PHE      & CE1  & 98     & LEU      & CD2  \\
22    & PHE      & CE1  & 98     & LEU      & CE1  \\
22    & PHE      & CE1  & 98     & LEU      & CD2  \\
24    & LEU      & CD1  & 129    & PHE      & CE1  \\
24    & LEU      & CD2  & 98     & LEU      & CD1  \\
24    & LEU      & CD2  & 98     & LEU      & CD2  \\
24    & LEU      & CD2  & 130    & GLY      & CA   \\
56    & ILE      & CG1  & 62     & CYS      & CB   \\
62    & CYS      & CB   & 139    & VAL      & CG1  \\
98    & LEU      & CD1  & 112    & PHE      & CB   \\
98    & LEU      & CD1  & 112    & PHE      & CD1  \\
98    & LEU      & CD1  & 129    & PHE      & CE1  \\
98    & LEU      & CD1  & 130    & GLY      & CA   \\
98    & LEU      & CD2  & 112    & PHE      & CB   \\
98    & LEU      & CD2  & 112    & PHE      & CD1  \\
98    & LEU      & CD2  & 129    & PHE      & CE1  \\ \hline
\end{tabular}
\caption{The $24$ atomic pairs in cyclophilin A for which both atoms are in core residues and NOE restraints have been measured. The atom types are reported in PDB format:  C$_{\alpha}$ (CA), C$_{\beta}$ (CB), C$_{\gamma}$ (CG), C$_{\delta}$ (CD), and C$_{\epsilon}$ (CE).}
\label{table:restraints}
\end{table}

\bibliography{MD_fluc}

\end{document}